\documentclass[11pt]{article}
\pdfoutput=1 
\usepackage{amssymb}
\usepackage{amsmath}
\usepackage{amstext}
\usepackage{graphicx,epsfig}
\usepackage{epsfig}
\usepackage{verbatim} 
\usepackage{fancybox}
\usepackage{color}
\usepackage{ulem}
\usepackage{enumitem}
\usepackage{youngtab}
\usepackage{subfigure}
\usepackage{bbm}
\usepackage{parskip}
\usepackage[numbers,sort&compress]{natbib}
\usepackage{ytableau}
\usepackage[utf8]{inputenc}

\usepackage{genyoungtabtikz}
\Yboxdim{13pt} 
\Ylinethick{.8pt}

\usepackage{tikz}
\usetikzlibrary{positioning, trees,decorations, decorations.markings, decorations.pathmorphing, arrows, shapes.geometric, snakes, patterns}
\pgfdeclaredecoration{complete sines}{initial}
{
\state{initial}[
width=+0pt,
next state=upsine,
persistent precomputation={\pgfmathsetmacro\matchinglength{
\pgfdecoratedinputsegmentlength / int(\pgfdecoratedinputsegmentlength/\pgfdecorationsegmentlength)}
\setlength{\pgfdecorationsegmentlength}{\matchinglength pt}
}] {}
\state{upsine}[width=\pgfdecorationsegmentlength,next state=downsine]{
\pgfpathsine{\pgfpoint{0.25\pgfdecorationsegmentlength}{0.5\pgfdecorationsegmentamplitude}}
\pgfpathcosine{\pgfpoint{0.25\pgfdecorationsegmentlength}{-0.5\pgfdecorationsegmentamplitude}}
}
\state{downsine}[width=\pgfdecorationsegmentlength,next state=upsine]{
\pgfpathsine{\pgfpoint{0.25\pgfdecorationsegmentlength}{-0.5\pgfdecorationsegmentamplitude}}
\pgfpathcosine{\pgfpoint{0.25\pgfdecorationsegmentlength}{0.5\pgfdecorationsegmentamplitude}}
}
\state{final}{}
}

\newcommand{\Comment}[1]{{}}
\definecolor{darkblue}{rgb}{0.15,0.35,0.55}
\definecolor{reddish}{rgb}{0.65, 0.2, 0.2}
\usepackage[linktocpage=true]{hyperref}
\hypersetup{
colorlinks=true,
citecolor=darkblue,
linkcolor=reddish,
urlcolor=darkblue,
pdfauthor={},
pdftitle={},
pdfsubject={}
}

\newcommand{\be}{\begin{equation}}
\newcommand{\ee}{\end{equation}}
\newcommand{\bea}{\begin{eqnarray}}
\newcommand{\eea}{\end{eqnarray}}
\newcommand{\beas}{\begin{eqnarray*}}
\newcommand{\eeas}{\end{eqnarray*}}
\newcommand{\nn}{\nonumber}

\definecolor{darkred}{rgb}{0.7,0.3,0.3}
\definecolor{darkgreen}{rgb}{0.2,0.7,0.3}
\definecolor{lightgreen}{rgb}{.816,.94,.753}
\definecolor{greyish}{rgb}{.8,.8,.8}
\definecolor{darkblue2}{rgb}{0.3,0.4,0.9}
\usepackage{pifont}
\usepackage{xcolor,colortbl}

\def\({\left(}
\def\){\right)}

\newcommand{\la}{\langle}
\newcommand{\ra}{\rangle}

\newcommand{\half}{\frac{1}{2}}

\def\gsim{ \lower .75ex \hbox{$\sim$} \llap{\raise .27ex \hbox{$>$}} }
\def\lsim{ \lower .75ex \hbox{$\sim$} \llap{\raise .27ex \hbox{$<$}} }

\def\xyma{\xymatrix@M.7em}
\def\xymas{\xymatrix@M.1em}

\title{}
\author{}

\numberwithin{equation}{section}

\begin{document}
\tikzset{
photon/.style={decorate, decoration={snake}, draw=magenta},
graviton/.style={decorate, decoration={snake}, draw=black},
sgal/.style={decorate , dashed, draw=black},
scalar/.style={decorate , draw=black},
mgraviton/.style={decorate, draw=black,
decoration={coil,amplitude=4.5pt, segment length=7pt}}
electron/.style={draw=blue, postaction={decorate},
decoration={markings,mark=at position .55 with {\arrow[draw=blue]{>}}}},
gluon/.style={decorate, draw=magenta,
decoration={coil,amplitude=4pt, segment length=5pt}} 
}
\renewcommand{\thefootnote}{\fnsymbol{footnote}}
~

\begin{center}
{\Large \bf Scale and Conformal Invariance on (A)dS \\
}
\end{center} 

\vspace{1truecm}
\thispagestyle{empty}
\centerline{\Large Kara Farnsworth,${}^{\rm a,}$\footnote{\href{mailto:kmfarnsworth@gmail.com}{\texttt{kmfarnsworth@gmail.com}}} Kurt Hinterbichler,${}^{\rm b,}$\footnote{\href{mailto:kurt.hinterbichler@case.edu} {\texttt{kurt.hinterbichler@case.edu}}} Ond\v{r}ej Hul\'{i}k,${}^{\rm c,}$\footnote{\href{mailto:ondra.hulik@gmail.com} {\texttt{ondra.hulik@gmail.com}}} }

\vspace{.5cm}

\centerline{{\it ${}^{\rm a}$D\'{e}partment de Physique Th\'{e}orique,}}
\centerline{{\it Universit\'{e} de Gen\`{e}ve, CH-1211 Gen\`{e}ve, Switzerland}} 
\vspace{.25cm}

\centerline{{\it ${}^{\rm b}$CERCA, Department of Physics,}}
\centerline{{\it Case Western Reserve University, 10900 Euclid Ave, Cleveland, OH 44106}} 
\vspace{.25cm}

\centerline{{\it ${}^{\rm c}$ Theoretische Natuurkunde, Vrije Universiteit Brussel,}}
\centerline{{\it Pleinlaan 2, B-1050 Brussels, Belgium}} 

\vspace{1cm}
\begin{abstract}
\noindent

We examine the question of scale versus conformal invariance on maximally symmetric curved backgrounds and study general 2-derivative conformally invariant free theories of vectors and tensors.   For spacetime dimension $D>4$, these conformal theories can be diagonalized into standard massive fields in which unbroken conformal symmetry non-trivially mixes components of different spins.   In $D=4$, the tensor case becomes a conformal theory mixing a partially massless spin-2 field with a massless spin-1 field.  For massless linearized gravity in $D = 4$, we confirm through direct calculation that correlation functions of gauge-invariant operators take the conformally invariant form, despite the absence of standard conformal symmetry at the level of the action. 

\end{abstract}
\newpage

\setcounter{tocdepth}{3}
\tableofcontents
\renewcommand*{\thefootnote}{\arabic{footnote}}
\setcounter{footnote}{0}

\newpage

\section{Introduction}

Quantum field theories (QFTs) are generically expected to arrive at renormalization group (RG) fixed points in their low energy limit (and high energy limit if the theory is UV complete). By definition, the physics of theories at RG fixed points is unchanged as the energy of the theory is scaled up or down. In flat space this is equivalent to the theory being invariant under a rigid scaling of the Cartesian coordinates,
\begin{align}
\delta x^\mu = \lambda x^\mu\, ,
\end{align} 
where $\lambda$ is a constant. When studied in detail, many of these theories are found to also be invariant under the entire conformal algebra, a larger algebra of transformations that includes the special conformal transformations,
\begin{align}
\delta x^\mu =  2(b\cdot x) x^\mu   - b^\mu x^2\,,
\end{align}
where $b^\mu$ is a constant vector.
Invariance under the full conformal algebra is very desirable; it is a highly constraining property, and there are numerous well developed techniques for understanding conformal theories. However it is unclear what properties a general theory must have in order to undergo this symmetry enhancement. There are proofs and strong arguments in  both two and four spacetime dimensions that scale invariant theories which are unitary, have a local stress energy tensor, and a discrete operator spectrum, are conformally invariant \cite{Polchinski:1987dy, Dymarsky:2013pqa,Dymarsky:2014zja,Yonekura:2014tha} as well as examples of non-unitary and non-local theories that are scale but not conformally invariant \cite{Riva:2005gd,Oz:2018yaz}. However in general dimension there are examples of theories with the same assumptions that do not have this symmetry enhancement \cite{Jackiw:2011vz,El-Showk:2011xbs,Mauri:2021ili} (see also \cite{Benedetti:2022zbb,Benedetti:2022ofj} for more general arguments from the algebraic QFT viewpoint). Recently, it was conjectured that shift symmetry may play a vital role, as it can explain the lack of enhancement to conformal invariance in known examples of interacting scale invariant theories \cite{Mauri:2021ili, Gimenez-Grau:2023lpz}, but this argument does not extend to free theories. Clearly more work is needed to understand the interplay between these two symmetries (see \cite{Nakayama:2013is} for a general overview of scale vs. conformal invariance). 

For theories with a local, gauge-invariant stress-energy tensor, the form of the trace of the stress-energy tensor and its correlation functions can be used to understand the presence of scale and/or conformal invariance. Here we wish to study this symmetry enhancement in theories which may not have such an operator.  The prime example of such a case is linearized gravity, the IR fixed point of quantum gravity. To understand the presence of scale and conformal invariance in these types of theories, we must look directly at the scale and conformal transformations of the correlation functions of the gauge-invariant operators present in the theory. 

We recently studied linearized gravity on flat space from this perspective \cite{Farnsworth:2021zgj}, as an example of a unitary theory with no stress tensor that is scale but not conformally invariant in $D>4$.  We found that, analogous to Maxwell theory \cite{Jackiw:2011vz,El-Showk:2011xbs}, in $D > 4$ the theory can be embedded into a non-unitary conformal field theory by adding specific gauge-fixing terms to the action. We also confirmed that, although there is no standard conformal invariance at the level of the action for linearized gravity in $D = 4$, the correlation functions of gauge-invariant operators, in this case the linearized Weyl tensor, do have the form dictated by conformal invariance. 

One of the aims in this paper is to extend the results of \cite{Farnsworth:2021zgj} to the other maximally symmetric spacetime backgrounds, namely de Sitter (dS) and anti-de Sitter (AdS) spacetimes. Conformal correlators on maximally symmetric backgrounds have previously been studied in e.g. \cite{Osborn:1999az,Hinterbichler:2015pta,Kikuchi:2019epb,Alvarez:2020pxc}, and conformally invariant equations have been classified \cite{Eastwood:1987ki,Shaynkman:2004vu}.  We begin by showing how there is no natural notion of a scale but not conformal theory on (A)dS, because there is no subalgebra of the conformal algebra of (A)dS that is strictly larger than the isometry algebra but strictly smaller than the conformal algebra, so there is no generator that one could call a scale transformation.   

We then go on to study all the possible bosonic single-field lower spin 2-derivative free conformal theories on (A)dS: scalar theories in section \ref{scalarsection}, vector theories in section \ref{vectorsection}, and spin-2 theories in section \ref{spin2section}.  For the scalar, it is well known that there is a specific mass value at which it becomes conformal.  For the vector, the massless theory is conformal for $D=4$, but we will see that for $D>4$ there is no mass value at which the Proca theory becomes conformal.  There is, however, a 2-derivative vector theory with non-canonical kinetic terms which is conformal for $D>4$ (and reduces to the massless case when $D=4$).  By diagonalizing this theory, we will see that it propagates a massive vector, with a particular mass analogous to the scalar's conformal mass, as well as a conformal scalar, and that the conformal symmetry acts in a non-trivial way which mixes the two fields while still acting linearly.  For the tensor, there is no massive Fierz-Pauli theory which is conformal.  The only conformal case is the massless tensor in $D=4$, which is not conformal in the standard way at the level of the action but whose correlators of gauge-invariant operators are in fact conformal.  For $D\geq 4$, there is a 2-derivative action with non-canonical kinetic terms which does have conformal symmetry.    For $D>4$, it diagonalizes into standard massive fields of spin $2$, $1$, and $0$, all with particular conformal masses generalizing the scalar's conformal mass.  The conformal symmetry mixes these fields into each other, while still acting linearly.  For $D=4$, this conformal action gets an enhanced scalar gauge symmetry, and the theory diagonalizes into a partially massless spin-2 field and a massless spin-1 field, again with conformal symmetry mixing them.  We will see that this conformal partially massless theory shares many properties with linearized conformal gravity.

\textbf{Conventions:} $D$ is the spacetime dimension, we use the mostly plus metric signature, and indices are (anti)symmetrized with weight one.  We use the curvature conventions of \cite{Carroll:2004st}.  Formulae are written for dS space where the radius is $1/H$, so that $R_{\mu\nu\rho\sigma}=H^2\left(g_{\mu\rho}g_{\nu\sigma}-g_{\mu\sigma}g_{\nu\rho}\right)$.  Results for AdS space of radius $L$ can be obtained via the replacement $H^2\rightarrow -1/L^2$.  Tensors are symmetrized and anti-symmetrized with unit weight ({\it e.g.} $t_{(\mu\nu)}=\half \left(t_{\mu\nu}+t_{\nu\mu}\right)$) and the notation $(\cdots)_T$ means we take the fully traceless and fully symmetric part of the enclosed indices.  Young tableaux are in the anti-symmetric convention.

\section{Conformal transformations, scale transformations and isometries\label{SvCAdS}}

The infinitesimal conformal symmetries of a metric $g_{\mu\nu}$ are those generated by conformal Killing vectors (CKV), vector fields $\xi^\mu$ that satisfy the conformal Killing equation
\be \nabla_\mu \xi_\nu+ \nabla_\nu \xi_\mu =2f g_{\mu\nu}\, ,\label{ckefee}\ee
where $f(x)$ is some scalar function.  Taking the trace of this equation implies $f= {1\over D} \nabla\cdot \xi$, and \eqref{ckefee} is equivalent to
\be  \nabla_\mu \xi_\nu+ \nabla_\nu \xi_\mu - {2\over D} \nabla\cdot \xi \, g_{\mu\nu} =0\,.\label{ckefee2} \ee
The set of vector fields satisfying these equations form a finite dimensional Lie algebra under the Lie bracket, which is the algebra of conformal symmetries of $g_{\mu\nu}$.  In spacetime dimensions $D>2$, the maximal dimension of this algebra is $(D+1)(D+2)/2$.

On the other hand, the isometries of a metric $g_{\mu\nu}$ are generated by Killing vectors, which satisfy the Killing equation $\nabla_\mu \xi_\nu+ \nabla_\nu \xi_\mu=0$.  These are the subset of conformal Killing vectors for which $f={1\over D}\nabla\cdot \xi=0$, and they form a subalgebra under the Lie bracket.  The maximal dimension of this algebra is $D(D+1)/2$.

We can define a scale transformation of a general spacetime as one for which the function $f$ on the right hand side of \eqref{ckefee} is a non-zero constant, i.e. $f= {1\over D}\nabla\cdot \xi=\textnormal{const.} \not=0$, so that the metric is simply rescaled by the transformation rather than distorted.  We will call the Killing vectors with this property scale Killing vectors (SKV), also called homothetic Killing vectors in the literature, see e.g. \cite{Stephani:2003tm}. These also form a subalgebra of the algebra of conformal Killing vectors under the Lie bracket, and contain as a subalgebra the algebra of isometries.  In summary, we have the inclusions 
\be {\rm isometries} \subseteq {\rm scale\ transformations} \subseteq {\rm conformal\ transformations}.\ee

\subsection{Flat space}

On flat space, $g_{\mu\nu}=\eta_{\mu\nu}$, a standard basis of conformal Killing vectors is the following,
\bea 
\begin{split}
P^\mu &= \partial^\mu\, ,  \\
J^{\mu\nu} &= x^\nu\partial^\mu-x^\mu\partial^\nu \, ,\\
D &= x^\mu\partial_\mu \, , \\
K^\mu &= 2x^\mu x^\nu\partial_\nu-x^2 \partial^\mu  \,.
\end{split}
 \label{flatckvse}
\eea
where $P_\mu$ and $J_{\mu\nu}$ generate translations and the Lorentz transformations respectively, which are the isometries of flat spacetime. $D$ and $K_\mu$ generate dilatations and the special conformal transformations respectively.  A generic conformal Killing vector field can be written as the linear combination,
\begin{align}
\xi = \xi^\mu \partial_\mu  = b^\mu K_\mu  + \lambda D +{1\over 2} \omega^{\mu \nu} J_{\mu\nu} + a^\mu P_\mu\,, 
\end{align}
with 
\begin{align}
\xi^\mu = 2(b\cdot x) x^\mu - b^\mu x^2  + \lambda x^\mu + \omega^\mu_{\ \nu} x^\nu + a^\mu\,.
\label{flatCKV}
\end{align}
Here we can see that for the isometries, parameterized by $a^\mu$ and $\omega^{\mu\nu}$, we indeed have $\partial \cdot \xi = 0$, and for the additional scale and special conformal transformations, parameterized by $\lambda$ and $b^\mu$, we have $\partial\cdot \xi \neq 0$ with $\partial \cdot f = \textnormal{constant}$ for the scale transformation.

The conformal Killing vectors satisfy the following non-zero commutations relations under the Lie bracket,
\bea
\begin{split}
 \left[D,P^\mu\right]&= -P^\mu \, , \\
 \left[D,K^\mu\right]&= K^\mu \,, \\
  \left[K^\mu,P^\nu\right]&=2(-\eta^{\mu\nu} D+J^{\mu\nu}) \, ,\\
   \left[J^{\mu\nu},K^\rho\right]&= \eta^{\rho\mu}K^\nu-\eta^{\rho\nu}K^\mu \, , \\
 \left[J^{\mu\nu},P^\rho\right] &= \eta^{\rho\mu} P^\nu-\eta^{\rho\nu} P^\mu\, , \\
\left[J^{\mu\nu},J^{\sigma\rho}\right]&=\eta^{\mu\sigma}J^{\nu\rho}-\eta^{\nu\sigma}J^{\mu\rho}+\eta^{\nu\rho}J^{\mu\sigma}-\eta^{\mu\rho}J^{\nu\sigma}\,.
\end{split}
 \label{ckgvcommtee}
\eea
This algebra is $\frak{so}(2,D)$.   
This can be seen more clearly by assembling the generators into a $(D+2)$-dimensional anti-symmetric matrix ${\mathbb J}^{AB}$ with $A,B=-1,0,1,\ldots,D$ as follows,
\be {\mathbb J}^{AB}=\left(\begin{array}{c|c|c}0 & D &  {1\over 2}\left({1\over m} P^\nu-m K^\nu\right) \\\hline -D & 0 & {1\over 2}\left({1\over m} P^\nu+m K^\nu\right) \\\hline -{1\over 2}\left({1\over m} P^\mu-m K^\mu\right) & -{1\over 2}\left({1\over m} P^\mu+m K^\mu\right) & J^{\mu\nu}\end{array}\right),\label{Jmatrixgenembsee}\ee
where $m$ is an arbitrary mass scale, upon which the commutation relations \eqref{ckgvcommtee} become
\be \left[{\mathbb J}^{AB},{\mathbb J}^{CD}\right]={\mathbb G}^{AC}{\mathbb J}^{BD}-{\mathbb G}^{BC}{\mathbb J}^{AD}+{\mathbb G}^{BD}{\mathbb J}^{AC}-{\mathbb G}^{AD}{\mathbb J}^{BC},\label{algesod2e}\ee
with
\be {\mathbb G}^{AB}=\left(\begin{array}{c|c|c}-1 &   &   \\\hline   & 1 &   \\\hline   &   & \eta^{\mu\nu}\end{array}\right),\ee
manifesting the structure of $\frak{so}(2,D)$.

The subalgebra of isometries, the conformal Killing vectors for which $\partial\cdot \xi=0$, are spanned by $J^{\mu\nu}$, $P^\mu$, forming the isometry algebra $\frak{iso}(1,D-1)$.  

The subalgebra of scale transformations is spanned by $J^{\mu\nu}$, $P^\mu$, $D$.  It is the isometry algebra along with the scale transformation $D$. 
This algebra lies strictly between the isometries and the conformal symmetries; it is strictly smaller than the full conformal algebra and strictly larger than the isometry subalgebra.  On flat space, it thus makes sense to ask if a theory can be scale but not conformally invariant.  Such a theory is invariant under all the conformal transformations except for those spanned by $K^\mu$, the special conformal generators.

\subsection{de Sitter space\label{dsscsec}}

Now consider de Sitter space in $D$ dimensions, dS$_D$.  The conformal Killing equations \eqref{ckefee2} are invariant under Weyl transformations, i.e. if $\xi^\mu$ is a CKV for the metric $g_{\mu\nu}$, it is also a CKV for the Weyl transformed metric $e^{2\sigma}g_{\mu\nu}$, for any scalar function $\sigma(x)$.  dS space is conformally related to flat space, which can be made explicit by using coordinates $x^\mu$ obtained by a stereographic projection of the dS hyperboloid, in which the metric takes the form
\be g_{\mu\nu}={1\over \left(1+{H^2\over 4} x^2\right)^2} \eta_{\mu\nu}\,,\label{dsmetrice}\ee
where $x^2\equiv \eta_{\mu\nu}x^\mu x^\nu$.   These coordinates are Riemann normal about the origin $x^\mu=0$, and the flat limit $H\rightarrow 0$ is manifest. Since the metric in these coordinates is a conformal factor times the flat metric, the expressions for the CKVs are exactly the same as the flat space ones in \eqref{flatckvse}, and so their generators satisfy the same algebra \eqref{ckgvcommtee} or \eqref{algesod2e}.  Thus the conformal algebra of dS$_D$ is $\frak{so}(2,D)$, the same as that of flat space.

To find the isometries of dS$_D$ in these coordinates, we need to find the subset of \eqref{flatckvse} that satisfies $\nabla_\mu \xi^\mu=0$, with $\nabla_\mu$ now the covariant derivative with respect to the metric \eqref{dsmetrice}.  Clearly the Lorentz transformations $J_{\mu\nu}$ are among these, since these are manifestly isometries of the metric \eqref{dsmetrice}.  Using the Weyl transformation rule $\nabla_\mu \xi^\mu=\partial_\mu \xi^\mu+{D\over \Omega} \xi^\mu\partial_\mu \Omega$ with $\Omega={1\over 1+{H^2\over 4}x^2}$, we can search for any other isometries, with the result that the linear combinations 
 \be \tilde P^\mu\equiv   P^\mu+{H^2\over 4} K^\mu\,,\ee
are the only other Killing vectors besides $J^{\mu\nu}$.  The commutators among the Killing vectors form the isometry algebra of dS$_D$,
\begin{align}
\begin{split}
 \left[J^{\mu\nu},J^{\sigma\rho}\right]&=\eta^{\mu\sigma}J^{\nu\rho}-\eta^{\nu\sigma}J^{\mu\rho}+\eta^{\nu\rho}J^{\mu\sigma}-\eta^{\mu\rho}J^{\nu\sigma}\,, \\
 \left[J^{\mu\nu},\tilde P^\rho\right] &= \eta^{\rho\mu} \tilde P^\nu-\eta^{\rho\nu} \tilde P^\mu\, , \\
 \left[\tilde P^\mu,\tilde P^\nu\right]&=H^2 J^{\mu\nu}\, .
 \end{split}
 \label{dsmanifestcorree}
\end{align}
In this form the flat limit $H\rightarrow 0$ manifestly reproduces the Poincar\'e algebra.
We can arrange the isometries into an $(D+1)$ dimensional anti-symmetric matrix ${\cal J}^{IJ}$ with $I,J=0,1,\cdots,D$ as follows,
\be {\cal J}^{IJ}=\left(\begin{array}{c|c}  0 & {1\over H}\tilde P^\mu \\\hline  - {1\over H}\tilde P^\mu & J^{\mu\nu}\end{array}\right),\label{Jmatrixgenembse2e}\ee
upon which the commutation relations \eqref{dsmanifestcorree} become
\be \left[{\cal J}^{I J},{\cal J}^{KL}\right]={\cal G}^{IK}{\cal J}^{JL}-{\cal G}^{JK}{\cal J}^{IL}+{\cal G}^{JL}{\cal J}^{IK}-{\cal G}^{IL}{\cal J}^{JK} ,\label{algesod2dSe}\ee
with
\be {\cal G}^{IJ}=\left(\begin{array}{c|c} 1 &   \\\hline      & \eta^{\mu\nu}\end{array}\right),\ee
showing that the isometry algebra of dS$_D$ is $\frak{so}(1,D)$.  We can also see that this is the subalgebra of the full conformal algebra consisting of the lower two rows and columns of \eqref{Jmatrixgenembsee}, upon choosing $m=H/2$.

We can add back the non-Killing combinations of conformal generators by packaging them into a $(D+1)$ dimensional vector ${\cal P}^I$ by defining\footnote{We can also express these in terms of embedding coordinates: let $X^I$, $I\in (0,\mu)$, be coordinates for a flat embedding space with metric $\eta_{AB}={\rm diag}(1,\eta_{\mu\nu})$.  Then the dS$_D$ hyperbola is the surface $\eta_{AB}X^AX^B={1\over H^2}$, parametrized in the coordinates \eqref{dsmetrice} by 
\be X^0= {1\over H}{-1+{H^2\over 4} x^2\over 1+{H^2\over 4} x^2},\ \ \ X^\mu= {x^\mu\over 1+{H^2\over 4} x^2}\,.\ee
Thinking of the $X^I$ as scalars on dS$_D$, the components of the conformal Killing vectors \eqref{nonkCKVe} can be written as gradients of these scalars,
\be \left({\cal P}^I \right)_\mu ={1\over H}\nabla_\mu X^I \,.\ee
These scalars also satisfy the equation
\be \nabla_\mu\nabla_\nu X^I=-H^2 g_{\mu\nu} X^I\,.\ee
 }
\be {\cal P}^0= D,\ \ \  {\cal P}^\mu={1\over H}\left( P^\mu-{H^2\over 4} K^\mu\right).\ \ \label{nonkCKVe} \ee
These are now the components ${\mathbb J}^{-1,I}$ of \eqref{Jmatrixgenembsee} and 
the full conformal commutation relations \eqref{ckgvcommtee}, \eqref{algesod2dSe} can be written as
\bea  
\begin{split} 
\left[{\cal J}^{I J},{\cal J}^{KL}\right]&={\cal G}^{IK}{\cal J}^{JL}-{\cal G}^{JK}{\cal J}^{IL}+{\cal G}^{JL}{\cal J}^{IK}-{\cal G}^{IL}{\cal J}^{JK} ,\\
\left[{\cal J}^{IJ},{\cal P}^K\right] &= {\cal G}^{KI} {\cal P}^J-{\cal G}^{KJ} {\cal P}^I\, , \\
\left[{\cal P}^I,{\cal P}^J\right] &=  - {\cal J}^{IJ} \, .
\end{split}
\label{algesofeuled2e}
 \eea
This manifests the dS isometry subalgebra of the full conformal algebra; the $\frak{so}(1,D)$ subalgebra is spanned by ${\cal J}^{IJ}$, and the non-isometries ${\cal P}^I$ transform in the vector representation of this $\frak{so}(1,D)$.

We can now address the question of whether there exists a scale but not conformal algebra on dS, i.e. whether the subalgebra of scale transformations is strictly smaller than the  conformal algebra $\frak{so}(2,D)$ but strictly larger than the isometry subalgebra $\frak{so}(1,D)$.  The answer is no: $\frak{so}(1,D)$ is a maximal subalgebra of $\frak{so}(2,D)$, so there is in fact no subalgebra which lies strictly between the isometries and the conformal transformations.   One way to see this is that since the ${\cal P}^I$ transform in the vector representation of $\frak{so}(1,D)$, and since this representation is irreducible, there is no way to pick only some of the ${\cal P}^I$'s and add them to $\frak{so}(1,D)$ without generating all the rest of them through commutators with the $\frak{so}(1,D)$ transformations.   It follows that there are no conformal Killing vectors for which $\nabla\cdot \xi$ is a non-zero constant.   In this sense, it is impossible even at the algebraic level to have a theory which is scale but not conformally invariant on dS.  An identical argument with some signs changed shows that the same is true of AdS.

\subsection{Transformations of fields}

We next recall how a rank $r$ tensor field $T_{\mu_1\cdots \mu_r}$ on a manifold with a general metric $g_{\mu\nu}$ transforms under conformal transformations.  To do this, it is convenient to first start with Weyl transformations, under which the metric and tensor field transform as
\be  g_{\mu\nu}\rightarrow e^{2\sigma}g_{\mu\nu},\ \ \ T_{\mu_1\cdots \mu_r}\rightarrow e^{-\Delta_W\sigma} T_{\mu_1\cdots \mu_r}\,,\label{fullweyleete}\ee
where $\sigma(x)$ is an arbitrary scalar function and $\Delta_W$ is a constant we call the Weyl weight of the tensor $T$.

Weyl transformations allow us to relate correlators of CFTs on conformally related spacetimes.  Given local primary operators ${\cal O}_1,{\cal O}_2,\ldots$ (potentially with tensor indices, which we suppress), with Weyl weights $\Delta^{({\cal O}_1)}_{W},\Delta^{({\cal O}_2)}_{W},\ldots$, the correlators on Weyl related spacetimes are related at separated points as\footnote{There can also be a Weyl anomaly that invalidates this formula, but it is local and so can only do so at coincident points.} 
\be \la {\cal O}_1(x_1){\cal O}_2(x_2)\cdots  \ra_{\Omega^2 g}=\Omega(x_1)^{-\Delta^{({\cal O}_1)}_{W}}\Omega(x_2)^{-\Delta^{({\cal O}_2)}_{W}}\cdots \la {\cal O}_1(x_1){\cal O}_2(x_2)\cdots  \ra_{g}   \, ,\label{weylrelatedcorree}\ee
with $\Omega(x) = e^{\sigma(x)}$. 

Infinitesimally, the Weyl transformation \eqref{fullweyleete} takes the form
\be \delta_W g_{\mu\nu}=2\sigma g_{\mu\nu},\ \ \ \delta T_{\mu_1\cdots \mu_r}=-\Delta_W\sigma \,T_{\mu_1\cdots \mu_r}\,.\ee
Under diffeomorphisms generated by an arbitrary vector $\xi^\mu$, the metric and tensor field transform via the Lie derivative,
\be \delta g_{\mu\nu}=-{\cal L}_\xi g_{\mu\nu}=-\left(\nabla_\mu\xi_\nu+\nabla_\nu\xi_\mu\right),\ \ \ \delta T_{\mu_1\cdots \mu_r}=-{\cal L}_\xi T_{\mu_1\cdots \mu_r}\,.\ee
The conformal transformations are precisely the combination of Weyl transformations and diffeomorphisms that leave the metric invariant, 
\be \left(\delta_W+\delta_D\right)g_{\mu\nu}=0\, ,\ee
which imposes $\nabla_\mu\xi_\nu+\nabla_\nu\xi_\mu=2\sigma g_{\mu\nu}$, meaning that $\xi^\mu$ is a conformal Killing vector and 
\be \sigma={1\over D}\nabla\cdot \xi\, .\ee
The tensor field then transforms as
\be \delta T_{\mu_1\cdots \mu_r}=-\left( {\cal L}_\xi +{\Delta_W\over D} \nabla\cdot \xi \right)T_{\mu_1\cdots \mu_r}\,.\label{gconfrftraee}\ee

On flat space, plugging in the conformal Killing vectors \eqref{flatckvse}, we get the transformation laws
\bea
\begin{split}
 \delta_{P^\mu}  &= -\partial^\mu \, , \\
\delta_{J^{\mu\nu} }&= x^\mu\partial^\nu-x^\nu\partial^\mu+{\cal J}^{\mu\nu} \, ,\\
\delta_ D &= -\left(x^\mu\partial_\mu+\Delta\right) \, ,\\
\delta_{K^\mu}  &=-2x^\mu x^\nu\partial_\nu+x^2 \partial^\mu -2 x^\mu\Delta-2x_\nu{\cal J}^{\mu\nu} \,,
\end{split}
\label{conformalffieldredrpse}
\eea
where $\left({\cal J}^{\mu\nu}\right)_{\mu_1\cdots\mu_r}^{\ \ \ \ \ \ \ \nu_1\cdots\nu_r}=\sum_{i=1}^r\delta_{\mu_{1}}^{\nu_{1}} \cdots \delta_{\mu_{i-1}}^{\nu_{i-1}} \left(2\delta_{\mu_i}^{[\mu} \eta^{\nu]\nu_i}\right)\delta_{\mu_{i+1}}^{\nu_{i+1}} \cdots \delta_{\mu_{r}}^{\nu_{r}} $ is the Lorentz generator on a rank $r$ tensor and
\be \Delta=\Delta_W+r\,.\label{deltatodwe}\ee
The transformation laws \eqref{conformalffieldredrpse} are those of a spinning conformal primary with conformal weight $\Delta$.

For dS space, we can evaluate \eqref{gconfrftraee} using the metric \eqref{dsmetrice}.  The Lie derivative does not depend on the metric and can be written using ordinary derivatives.  Using the Weyl transformation rule $\nabla_\mu \xi^\mu=\partial_\mu \xi^\mu+{D\over \Omega} \xi^\mu\partial_\mu \Omega$ with $\Omega={1\over 1+{H^2\over 4} x^2}$, we evaluate the rest in terms of ordinary derivatives giving the rule
\be \delta T_{\mu_1\cdots \mu_r}=-\left[ {\cal L}_\xi +{\Delta-r\over D} \left( \partial\cdot \xi -{D\over 2} { H^2 \xi\cdot x\over 1+{H^2\over 4} x^2}\right) \right]T_{\mu_1\cdots \mu_r}\,,\label{gconfrftraedsse}\ee
where we have also used \eqref{deltatodwe}.  When $\xi^\mu$ is one of the dS isometries, this reduces to only the Lie derivative term.  When $H\rightarrow 0$, it reduces to the flat space conformal transformations.

\section{Scalars\label{scalarsection}}

The Lagrangian for a canonical free scalar $\phi$ of mass $m$ on dS$_D$ is
\be {1\over \sqrt{-g}}{\cal L}=-{1\over 2}\nabla_\mu \phi \nabla^\mu \phi -{1\over 2} m^2\phi^2\,.\label{scalaracele}\ee

\subsection{Conformal symmetry}

We know that the Lagrangian \eqref{scalaracele} should be conformally invariant for a particular value of the mass where the mass term corresponds to the conformal coupling to the Ricci scalar, which is a constant for dS.  
The conformal transformation of the scalar is \eqref{gconfrftraedsse} with $r=0$,
\be \delta \phi=-\left[\xi^\mu\partial_\mu +{\Delta\over D} \left( \partial\cdot \xi -{D\over 2} { H^2 \xi\cdot x\over 1+{H^2\over 4} x^2}\right) \right]\phi \,.\label{gconfrftraedsse2}\ee
Demanding that \eqref{scalaracele} be invariant up to a total derivative under \eqref{gconfrftraedsse2} fixes\footnote{\label{Dfootnote}Note that in checking the conformal symmetry, it suffices to check it for only one of the non-isometry conformal generators, for example the generator $D$ for which 
\be \xi^\mu=\lambda x^\mu,\ \ \  \delta \phi=-\lambda \left[x^\mu\partial_\mu +{\Delta}  -{1\over 2}{\Delta\, H^2 x^2\over 1+{H^2\over 4} x^2}  \right]\phi \,.\ee
  The rest are then guaranteed to be symmetries because of the irreducibility of ${\cal P}^I$ as discussed at the end of section \ref{dsscsec}.}
\be m^2={D(D-2)\over 4} H^2\,, \ \ \ \Delta={D\over 2}-1\,,\label{scalarconfvaluee}\ee
which is the well-known conformal mass value and canonical dimension for a free scalar.  This also shows that there are no other values of $m^2$ and $\Delta$ for which the action \eqref{scalaracele} is conformally invariant with $\phi$ transforming in the standard way.  

This conformal symmetry can be seen as a consequence of the existence of the Weyl invariant Lagrangian where the field has Weyl weight $\Delta_W=\Delta={D\over 2}-1$,
\begin{align}
 {1\over \sqrt{-g}}{\cal L}_{{\rm Weyl},\phi} = -\frac{1}{2}\nabla_\mu \phi \nabla^\mu \phi - \frac{(D-2)}{8(D-1)} R \phi^2\,,
\end{align}
which reduces to \eqref{scalaracele} on dS where the Ricci scalar $R = D(D-1)H^2$.

\subsection{Correlator}

The conformal mass value for \eqref{scalaracele} can also be seen directly from correlation functions of $\phi$.  Consider for simplicity the Euclidean dS correlator $\la \phi(x)\phi(x')\ra$ (i.e. the correlator on the sphere of radius $1/H$).  The field
 satisfies the free Euclidean equation of motion
\be \left(\nabla^2-m^2\right)\phi=0\,,\label{correlatoreom} \ee
and so the correlator must satisfy this at separated points.

By dS symmetry, the correlator must be a function of the geodesic distance $\mu$ between $x$ and $x'$, which we express in terms of the variable $Z$ defined as $Z =\cos\left(H\mu\right)$ (see Appendix \eqref{appendixbi-tensor}),
\be \la \phi(x)\phi(x')\ra=G(Z).\ee
 On the sphere, $Z\in [-1,1]$ and $\mu\in [0,\pi/H]$, with $Z=1$, $\mu=0$ the coincident point and $Z=-1$, $\mu=\pi/H$ the antipodal point.  In terms of the variable $Z$, the equation of motion \eqref{correlatoreom} at separated points becomes
 \be (Z^2 - 1) G''(Z) + D Z G'(Z)+ {m^2\over H^2} G(Z) =0\,. \label{scZee}\ee
The two boundary conditions that fix the solution to this second order differential equation are: 1) the solution should be regular at the antipodal point, and 2) the coincident-point singularity should be the same as that of the flat space correlator,
\be G(Z(\mu))\underset{\mu\rightarrow 0}{\Rightarrow} {\Gamma\left({D/ 2}-1\right)\over 4\,\pi^{D/2} }{1\over \mu^{D-2}}\,.\ee
Enforcing these boundary conditions, the solution to \eqref{scZee} is
\be G(Z)={H^{D-2}\Gamma(\delta_+)\Gamma(\delta_-)\over 2^D \pi^{D/2} \Gamma\left(D/2\right)}\, {_{2}F_1} \left( \delta_-,\delta_+;{D\over 2};{Z+1\over 2}\right),\label{genscalarproze} \ee
where $\delta_\pm$  are the scalar ``dual conformal dimensions'' corresponding to the mass obtained via the (A)dS/CFT mass formula \eqref{gensadscctm},
\be \textnormal{scalar:}\quad{} \delta_\pm \equiv {d\over 2}\pm\sqrt{{d^2\over 4}-{m^2\over H^2}},\ \ \ d\equiv D-1\,.\ee

As a function of $m^2$, \eqref{genscalarproze} is well defined except for poles coming from $\Gamma(\delta_-)$ at the values 
\be m^2=-k (k + D -1)H^2\, ,\ \ \ k=0,1,2,\ldots\ \ ,\label{shiftsymmass} \ee
for which $\delta_-=-k$. These are the values at which the scalar acquires enhanced Galileon-like shift symmetries \cite{Bonifacio:2018zex,Bonifacio:2021mrf}, the first case being the massless scalar $k=0$ which has the simple shift symmetry $\phi\rightarrow \phi+\textnormal{const.}$  At these values, the shift symmetries should be gauged \cite{Folacci:1992xc}, and correlators of the shift invariant operators will be finite. 

At the conformal value of the mass \eqref{scalarconfvaluee}, we have $\delta_\pm={d\pm 1\over 2}$ and \eqref{genscalarproze} simplifies to
\be G(Z)= {H^{D-2}\Gamma(D/2)\over(D-2)  (2\pi)^{D/2}}{1\over \left( 1-Z\right)^{D/ 2-1}}\,.\label{confscorre}\ee
(When $D=2$ we see a pole because the conformal value in this case, $m^2 = 0$, coincides with the $k=0$ shift symmetric value.)
In the coordinates $x^\mu$ with the metric \eqref{dsmetrice}, the geodesic distance between points $x^\mu$ and $x'^\mu$ is given by \eqref{conformalepforZe},
and \eqref{confscorre} becomes
\be G(x,x')= \Omega(x)^{-\Delta} \Omega(x')^{-\Delta}\left[ {\Gamma\left(\Delta\right)\over 4\, \pi^{\Delta+1} }{1\over |x-x'|^{2\Delta}}\right], \label{sqbreese}
\ee
with
\be
 \Omega(x)={1\over 1+{H^2\over 4} x^2},\ \ \Delta={D\over 2}-1\, .\label{confscorrxe}\ee
The factor in square brackets in \eqref{sqbreese} is the correlator of a free massless scalar on flat space, which has the conformally invariant structure $\sim { |x-x'|^{-2\Delta}}$.  This demonstrates the formula \eqref{weylrelatedcorree} and \eqref{deltatodwe}.

\section{Vectors\label{vectorsection}}

A canonical free vector field $A_\mu$ of mass $m$ on dS$_D$ is described by the Proca Lagrangian,
\be {1\over \sqrt{-g}}{\cal L}=-{1\over 4}F_{\mu\nu}^2-{1\over 2} m^2A^2\,,\label{vescalaracele}\ee
where $F_{\mu\nu}\equiv\nabla_\mu A_\nu-\nabla_\nu A_\mu$.

For $m=0$ we have the usual electromagnetic gauge symmetry
\be \delta A_{\mu}=\nabla_\mu\Lambda,\label{emgaugesyme}\ee
with scalar gauge parameter $\Lambda$.

In what follows we will restrict to $D>2$, ignoring some possibly interesting subtleties that occur when $D=2$, where the massive vector is dual to a scalar.

\subsection{Conformal symmetry}

In the coordinates \eqref{dsmetrice}, under a conformal transformation the vector should transform as \eqref{gconfrftraedsse} with $r=1$,
\be \delta A_{\mu}=-\left[ \xi^\nu \partial_\nu A_\mu +\partial_\mu\xi^\nu A_\nu   +{\Delta-1\over D} \left( \partial\cdot \xi -{D\over 2} { H^2 \xi\cdot x\over 1+{H^2\over 4}x^2}\right)A_\mu \right].\label{gconfrftraedvsse}\ee
A brute force calculation demanding that the Lagrangian in \eqref{vescalaracele} be invariant up to a total derivative under \eqref{gconfrftraedvsse} with $\xi^\mu=\lambda x^\mu$ (see footnote \ref{Dfootnote}) shows that the only case that is conformally invariant is
\be m^2=0\, , \ \ \ \Delta=1\, ,\ \ \ D=4\,.\label{maxwellsole}\ee
Thus only the massless vector has a conformally invariant action, and only in $D=4$, where it has the scaling dimension $\Delta=1$.  In flat space, the massless vector is scale invariant but not conformally invariant in $D\not=4$ \cite{Jackiw:2011vz,El-Showk:2011xbs}; there is no analog of this in (A)dS as per the discussion in section \ref{dsscsec}.

There is a conformal theory away from $D=4$ if we allow for the most general Lagrangian with up to two derivatives.  Consider the Lagrangian
\be {1\over \sqrt{-g}}{\cal L}=-{1\over 4}F_{\mu\nu}^2-{\xi\over 2} (\nabla\cdot A)^2-{1\over 2} m^2A^2\,,\label{vescalaracelegke}\ee 
where $\xi$ parametrizes the deviation from the Maxwell kinetic structure.  Demanding that this Lagrangian be invariant up to a total derivative under \eqref{gconfrftraedvsse} yields a unique solution\footnote{In the ansatz \eqref{vescalaracelegke} we have implicitly assumed that the leading structure $(\nabla_\mu A_\nu)^2$ does not vanish.  Relaxing this assumption does not yield any other non-trivial solutions. }
\be \xi={D-4\over D},\ \ \ m^2={(D-4)(D-2)\over 4}H^2,\ \ \ \Delta={D\over 2}-1\,,\label{mvsole}\ee
giving the Lagrangian
\be {1\over \sqrt{-g}}{\cal L}_A=-{1\over 4}F_{\mu\nu}^2-{(D-4)\over 2D} (\nabla\cdot A)^2-{(D-4)(D-2)\over 8}H^2A^2\,.\label{vescalaracelegke2}\ee
This coincides with the massless vector solution \eqref{maxwellsole} when $D=4$.  In $D>4$, this is an (A)dS version of the $\kappa=1$, $s=1$ ``special'' conformal field in the classification of \cite{Metsaev:2016oic}. 
It was found earlier in \cite{Deser:1983mm}, and discussed further along with higher spin generalizations in \cite{Iorio:1996ad,Erdmenger:1997wy,Vasiliev:2009ck,Chekmenev:2020lkb,Asorey:2021rwv}.  It also appears in tractor constructions of conformal fields \cite{Gover:2008pt,Gover:2008sw}.  

The theory \eqref{vescalaracelegke2} comes from the following Weyl invariant Lagrangian, where $A_\mu$ transforms with Weyl weight $\Delta_W=\Delta-1={D\over 2}-2$,
\be   {1\over \sqrt{-g}}{\cal L}_{{\rm Weyl},A}=-{1\over 4}F_{\mu\nu}^2-{(D-4)\over 2D} (\nabla\cdot A)^2+{(D-4)\over 2(D-2)} A^\mu A^\nu \left( R_{\mu\nu}-{D\over 4(D-1)} R g_{\mu\nu}\right). \label{vescalaracelegke2ewe}\ee 
This reduces to \eqref{vescalaracelegke2} when restricted to dS space.

For $D\not=4$ the conformal model \eqref{vescalaracelegke2} is not purely a massive vector because the altered kinetic term introduces new degrees of freedom.  (Nor can it be considered a gauge-fixed massless vector: the mass term is non-vanishing and it propagates massive modes.)  One way to see this is to look at the equations of motion from \eqref{vescalaracelegke2} which read 
\be\left(  \nabla^2 -{(D^2-2D+4)\over 4}H^2 \right) A_\mu-{4\over D} \nabla_\mu \left (\nabla\cdot A\right) =0\, .\label{confvecteqe}\ee
Taking a divergence, we find
\be {(D-4)\over D}\left( \nabla^2 -{D(D-2)\over 4}H^2\right) \nabla\cdot A=0\, .\label{Aconstraint}\ee
The analogous divergence on the equation of motion for a massive vector would have yielded a constraint $\nabla\cdot A=0$.  Here we instead find a wave equation for the scalar quantity $\nabla\cdot A$.  In fact, it is the wave equation for a conformal scalar with the mass \eqref{scalarconfvaluee}, so for $D\not=4$ we can expect a propagating conformal scalar mode in addition to a vector mode.

We can explicitly separate the extra physical mode at the level of the Lagrangian as follows.  Starting with \eqref{vescalaracelegke2},
we introduce an auxiliary scalar $\phi$ through
\be  {1\over \sqrt{-g}}{\cal L}=-{1\over 4}F_{\mu\nu}^2-{(D-4)(D-2)\over 8}H^2A^2+\phi \nabla\cdot A + {D\over 2(D-4)}\phi^2\,.\label{vescalaracelegke3}\ee 
The $\phi$ equations of motion give
\be \phi=-{(D-4)\over D} \nabla\cdot A\,,\ee
which upon plugging into \eqref{vescalaracelegke3} recovers \eqref{vescalaracelegke2}.  We can then diagonalize \eqref{vescalaracelegke3}  by the field redefinition
\be A_\mu\rightarrow \tilde{A}_\mu -{4\over H^2(D-4)(D-2)}\nabla_\mu\phi\, , \label{Atoaphiedfre}\ee
after which we have
\be  {1\over \sqrt{-g}}{\cal L}=-{1\over 4}{\tilde{F}}_{\mu\nu}^2-{1\over 2}\tilde{m}^2_{\tilde A}{\tilde{A}}^2 -{1\over \tilde{m}^2_{\tilde A}}  \left[-{1\over 2}(\nabla\phi)^2-  {1\over2}m_\phi^2\phi^2\right]\, ,\label{Aaplage}\ee
where 
\be \tilde{m}^2_{\tilde A}={(D-4)(D-2)\over 4}H^2,\ \  m_\phi^2 ={D(D-2)\over 4}H^2\,.\label{scandvecmee}\ee
We now see that the dynamical degrees of freedom are those of a massive vector and a massive scalar, with masses given by \eqref{scandvecmee}.
The scalar mass is the conformal mass \eqref{scalarconfvaluee}.  The vector mass can be considered a generalization of the scalar's conformal mass: {note that $\tilde{A}$ and $\phi$ both have the same dual CFT dimension $\delta_\pm={d\pm 1\over 2}$, where $d \equiv D-1$.} 
 For $D>4$ the vector and scalar kinetic terms have opposite signs on dS and the same signs on AdS.  However, the sign of the vector mass indicates that the vector field is non-unitary on AdS and healthy on dS.  For $D=3$, the same statements are true with dS and AdS reversed. Thus for $D\not=4$, the combined set of fields is always non-unitary on both dS and AdS. 

The Lagrangian in \eqref{Aaplage} is conformally invariant since it was arrived at by field redefinitions from the conformally invariant Lagrangian \eqref{vescalaracelegke2}.  However, because of the diagonalizing field redefinition \eqref{Atoaphiedfre}, in terms of the fields $\tilde{A}_\mu$, $\phi$, the conformal symmetry acts in a complicated way, mixing the two fields together (though it is still linearly realized).  The two fields $\tilde{A}_\mu$ and $\phi$, each a separate irreducible representation from the point of view of the dS algebra, join together into a larger irreducible representation of the larger conformal algebra (see \cite{Penedones:2023uqc} for more on how conformal representations break up into dS representations).

\subsection{Correlator}

We now turn to explicitly computing the correlators for the conformal vector theories, in order to confirm that they take the unique conformally invariant form.
In $D\neq 4$, the theory \eqref{vescalaracelegke2} has no gauge invariance and is conformally invariant at the level of the action, and thus we should be able to see conformal invariance directly by studying the correlators of the basic field $A_\mu$.  In order to deal with the non-canonical structure of these cases, we will first compute the correlators of the component massive fields in the diagonalized Lagrangian \eqref{Aaplage}, and then use these to form the correlator of the original field through the field redefinition \eqref{Atoaphiedfre}.   In $D = 4$, conformal invariance requires $m^2 = 0$ and the action \eqref{vescalaracele} has the Maxwell gauge symmetry $\delta A_\mu = \nabla_\mu \Lambda$. In this case, we must look at the two-point function of the basic gauge-invariant operator, namely the field strength $F_{\mu\nu} = \nabla_\mu A_\nu - \nabla_\nu A_\mu$. This serves as a warm up for the spin-2 case, where the link between conformal invariance of the correlators and standard conformal invariance of the action is broken.

Consider first the generic massive Proca theory \eqref{vescalaracele}.  We will need the Euclidean dS two-point function $\langle A_\mu (x) A_{\nu'} (x')\rangle$.
Maximal symmetry implies that this correlator has the following structure \cite{Allen:1985wd} (see also \cite{Narain:2014oja,Belokogne:2016dvd}), 
\begin{align}
 \langle A_\mu (x) A_{\nu'} (x')\rangle= f_1(Z) g_{\mu\nu'} + f_2(Z) n_\mu n_{\nu'}\,,
\label{veccorrgen}
\end{align}
where $g_{\mu\nu'}(x,x')$ is the parallel propagator along the geodesic between $x$ and $x'$, and $n_\mu(x,x')$ and $n_{\nu'}(x,x')$ are the unit tangent vectors to the geodesic at the points $x$ and $x'$ (this notation follows \cite{Allen:1985wd}, see Appendix \ref{appendixbi-tensor} for a review and definitions and expressions in our particular coordinate system).  Here $f_1(Z)$ and $f_2(Z)$ are arbitrary functions of the geodesic distance $\mu$ expressed through the variable $Z\equiv\cos(H\mu)$.

The equations of motion for the massive vector ${A}_\mu$ are equivalent to
\begin{align}
\left(\nabla^2 - (D-1)H^2  -m^2\right) {{A}}_\mu = 0\, ,\ \ \ \nabla_\mu A^\mu=0\,.\label{veceomee}
\end{align}
The two-point function \eqref{veccorrgen} must obey these equations of motion at separated points.

Plugging the structure \eqref{veccorrgen} into the equations of motion \eqref{veceomee} gives
\begin{align}
\begin{split}
f_1(Z)&=\frac{ (Z^2-1)}{(D-1)}G'(Z)+ZG(Z)\, \\ 
f_2(Z) &= \frac{ (Z^2-1)}{(D-1)}G'(Z)+(Z-1)G(Z)\, 
\end{split}
\label{abexprve}
\end{align}

where $G(Z)$ satisfies the differential equation
\begin{align}
(Z^2-1)G''(Z)+(D+2)ZG'(Z) + \left(2(D-1) + \frac{m^2}{H^2}\right)G(Z) = 0\, .
\label{diffeqX}
\end{align}

The solution to \eqref{diffeqX}, with the boundary conditions for Euclidean dS (regularity at the antipode and matching to flat space behavior at the coincident point), is
\begin{align}
G(Z) = {H^{D}\over m^2}\frac{(1-D) \Gamma(\delta_++1) \Gamma(\delta_-+1)}{2^{D+1} \pi^{D/2} \Gamma\left(D/2 + 1\right)}   \, {}_2 F_1\left(\delta_-+1, \delta_++1; \frac{D+2}{2}; \frac{Z+1}{2}\right),
\end{align}
with
\begin{align}
\textnormal{vector:}\quad{} \delta_{\pm}
&= \frac{d}{2}  \pm \sqrt{\frac{(d-2)^2}{4} - \frac{m^2}{H^2}},\quad{} d \equiv D-1.
\end{align}
The expressions for $f_1(Z)$ and $f_2(Z)$ from \eqref{abexprve} can then be written as
\begin{align}
\begin{split}
f_1(Z)=&\ (1-Z^2)\frac{\Gamma(\delta_++2) \Gamma(\delta_-+2)}{2^{D} \pi^{D/2} \Gamma\left(D/2\right)}  \frac{H^{D}}{m^2}\,  {}_2 F_1\left(\delta_-+2, \delta_++2; \frac{D+4}{2}; \frac{Z+1}{2}\right)\\
&+Z\frac{(1-D) \Gamma(\delta_++1) \Gamma(\delta_-+1)}{2^{D+1} \pi^{D/2} \Gamma\left(D/2+ 1\right)} \frac{H^{D}}{m^2}\,   {}_2 F_1\left(\delta_-+1, \delta_++1; \frac{D+2}{2}; \frac{Z+1}{2}\right),\\
f_2(Z) =&\ (1-Z^2)\frac{\Gamma(\delta_++2) \Gamma(\delta_-+2)}{2^{D} \pi^{D/2} \Gamma\left(D/2 \right)} \frac{H^{D}}{m^2}\,  {}_2 F_1\left(\delta_-+2, \delta_++2; \frac{D+4}{2}; \frac{Z+1}{2}\right)\\
&+(Z-1)\frac{(1-D) \Gamma(\delta_++1) \Gamma(\delta_-+1)}{2^{D+1} \pi^{D/2} \Gamma\left(D/2+ 1\right)} \frac{H^{D}}{m^2}\,   {}_2 F_1\left(\delta_-+1, \delta_++1; \frac{D+2}{2}; \frac{Z+1}{2}\right).
\end{split}
\label{MassiveAcoeff}
\end{align}

The correlator becomes singular at the massless value $m^2=0$.  This is where the theory becomes gauge invariant under \eqref{emgaugesyme}.  As we will see shortly, this singularity in the propagator will cancel in correlators of gauge invariant observables.

The correlator also becomes singular, via the poles of $\Gamma(\delta_-+1)$, at the mass values
\be m^2  = -(k+2)(k+D-1)H^2\, ,\ \ \ k=0,1,2,\ldots\,  , \ee
for which $\delta_+=d+k+1$, $\delta_-=-k-1$.  These are the values at which the massive vector acquires enhanced shift symmetries \cite{Bonifacio:2018zex,Bonifacio:2019hrj}.  At these values, the shift symmetries should be gauged and correlators of the shift invariant operators will be finite.

\subsubsection{Conformal theory in $D \neq 4$} 
For $D \neq 4$, we saw that the conformally invariant non-canonical vector Lagrangian \eqref{vescalaracelegke2} for the field $A_\mu$ can be rewritten as a massive vector $\tilde A_\mu$ plus a massive scalar $\phi$ as in \eqref{Aaplage}, with masses \eqref{scandvecmee}.
To find the correlators of the original vector field $A$, we can therefore compute the correlators of the Proca vector $\tilde{A}_\mu$ and the scalar $\phi$, and then combine them using \eqref{Atoaphiedfre}.  

The scalar part of the action \eqref{Aaplage} is proportional to the action \eqref{scalaracele} for a massive scalar with the conformal mass value in \eqref{scalarconfvaluee}, so the two-point function in this case is just the two-point function \eqref{confscorre} with the appropriate normalization,
\be 
\langle \phi(x) \phi(x')\rangle =-\tilde{m}^2_{\tilde A} {H^{D-2}\Gamma(D/2)\over (D-2) (2\pi)^{D/2}}{1\over \left( 1-Z\right)^{{D/ 2}-1}}\, . 
\ee

The vector part of the action \eqref{Aaplage} is proportional to the action \eqref{vescalaracele} for a massive vector with the conformal mass value $\tilde{m}^2_{\tilde A}$ in \eqref{scandvecmee}, so the two-point function in this case is just the massive vector two-point function \eqref{veccorrgen}, \eqref{MassiveAcoeff} with this mass value, which simplifies dramatically to
\begin{align} 
&\langle \tilde A_\mu(x) \tilde A_{\nu'}(x')\rangle=\label{Atilde2pt} \\ &\quad  
 \frac{ H^{D-2}\Gamma(D/2)}{(D-2)(D-4)(2 \pi)^{D/2}}\frac{1}{(1-Z)^{D/2}} \bigg[\left[ D- (D-2)Z\right] g_{\mu\nu'}+  \left[ (3D-2)- (D-2)Z\right] n_\mu n_{\nu'}\bigg].\nn
\end{align}
Now, using the relation \eqref{Atoaphiedfre}, which tells us
\be {A}_\mu= \tilde A_\mu -{1\over \tilde{m}^2_{\tilde A}} \nabla_\mu\phi\, ,\ee
we can form the correlator of the original $A_\mu$,
\begin{align}
\begin{split}
 \langle  A_\mu(x)  A_{\nu'}(x')\rangle &= \langle \tilde A_\mu(x) \tilde A_{\nu'}(x')\rangle +{1\over \tilde{m}^4_{\tilde A}} \nabla_\mu \nabla_{\nu'}  \langle\phi(x) \phi(x')\rangle\, . \label{combinationsve}
\end{split}
\end{align}
The derivatives acting on the scalar two-point function can be computed using the formulas in Appendix \ref{appendixbi-tensor}, 
\begin{align}
& \frac{1}{\tilde{m}^4_{\tilde A}} \nabla_\mu \nabla_{\nu'}  \langle\phi(x) \phi(x')\rangle =\label{Ascalar2pt} \\ &\quad -\frac{H^{D-2}\Gamma(D/2)}{(D-2)(D-4)(2\pi)^{D/2}}\frac{1}{(1-Z)^{D/2}} \left[ 2g_{\mu\nu'} + \left(D+2+(D-2)Z \right)n_\mu n_{\nu'}\right].\nn
\end{align}

Combining \eqref{Atilde2pt} and \eqref{Ascalar2pt} as in \eqref{combinationsve}, we arrive at a simple form for the two point function of our original vector field $A_\mu$,
\begin{align}
\begin{split}
\langle A_\mu(x) A_{\nu'}(x')\rangle = \frac{ H^{D-2}\Gamma(D/2)}{(D-4) (2 \pi)^{D/2}}\frac{1}{(1-Z)^{D/2-1}} \left(g_{\mu\nu'}+  2 n_\mu n_{\nu'}\right).
\end{split}
\label{confAcorr}
\end{align}
Writing this in our conformally flat coordinate system \eqref{dsmetrice} using the expressions in Appendix \ref{appendixbi-tensor}, we have,
\begin{align}
\langle A_\mu(x) A_{\nu'}(x')\rangle =&\  \Omega(x)^{-\Delta +1}\Omega(x')^{-\Delta+1}\left[ \frac{2\Delta}{(D-4)}\frac{\Gamma(\Delta)}{4\pi^{\Delta+1}}\frac{1}{|x-x'|^{2\Delta}}I_{\mu\nu'}(x,x')\right],\label{vecdg4finalce}
\end{align}
where
\begin{align}
\Omega(x) = \frac{1}{\left(1+\frac{H^2}{4} x^2\right)},\quad{} \Delta = \frac{D}{2} - 1\,, \label{vecdeltavaluee}
\end{align}
and 
\begin{align}
I_{\mu\nu'}(x,x') \equiv \eta_{\mu\nu'}- 2\frac{(x-x')_\mu (x-x')_{\nu'}}{(x-x')^2} \,, \label{confstruct}
\end{align}
is the conformally invariant tensor structure of \cite{Osborn:1993cr} in flat space. 
The term in square brackets in \eqref{vecdg4finalce} has the structure of a two-point function of a conformal primary of spin $1$ and dimension $\Delta$ in flat space. Out front we see the Weyl factors for the transformation of this two-point function from flat space to dS, with Weyl weight $\Delta_W = \Delta - 1$ for the rank-1 field, confirming \eqref{weylrelatedcorree} and \eqref{deltatodwe} and demonstrating that this non-canonical vector theory is conformal on dS$_D$.

\subsubsection{Maxwell theory in $D = 4$}

The two-point function for the massless vector in maximally symmetric spacetimes was calculated explicitly in \cite{Allen:1985wd}. In this section we reproduce the calculation in a different way, showing that the two-point function of the field strength has the expected form of a conformal primary in $D = 4$. Unlike \cite{Allen:1985wd}, we start from the massive vector in general $D$ and carefully take the massless limit, isolating the singularity that appears due to the gauge symmetry \eqref{emgaugesyme}. Taking the appropriate gauge-invariant linear combination of derivatives for the two-point function, the singularity cancels and we get a finite conformally invariant correlator for the gauge-invariant field strength.

Starting with the expressions \eqref{MassiveAcoeff} for the coefficients in the massive vector two point function \eqref{veccorrgen}, setting $m^2 = \epsilon H^2$ and expanding in $\epsilon$, we have\footnote{Here and below, we found the Mathematica package HypExp \cite{Huber:2005yg} useful for expanding the hypergeometric functions.} 
\begin{align}
\begin{split}
f_1(Z)&=\frac{H^2}{8\pi^2(Z-1)^2}\bigg[\frac{ 2-Z}{\epsilon }+\frac{\left(Z^3-3 Z+2\right) \log \left(\frac{1-Z}{2}\right)-Z^2+1}{(Z+1)^2}+\mathcal{O}(\epsilon)\bigg] ,\\
f_2(Z) &=\frac{H^2}{8\pi^2(Z-1)^2}\bigg[\frac{5-Z}{\epsilon }+\frac{(Z-1)^3 \log \left(\frac{1-Z}{2}\right)+2(Z^2-1)}{(Z+1)^2}+\mathcal{O}(\epsilon)\bigg].
\end{split}
\end{align}
The divergence at $m^2=0$ now appears as a simple pole in $\epsilon$.

Taking derivatives and antisymmetrizing to form the two-point function for the field strength $F_{\mu\nu} = \nabla_\mu A_\nu - \nabla_\nu A_\mu$ gives
\begin{align}
&\langle F_{\mu\nu}(x)F_{\mu'\nu'}(x')\rangle =\\ &\quad  
 \frac{H^4}{4 \pi ^2 (Z-1)^2}\left[\left( g_{\mu\mu'}+2n_\mu n_{\mu'}\right)\left(g_{\nu\nu'}+2n_\nu n_{\nu'}\right)- \left(g_{\nu\mu'}+2n_\nu n_{\mu'}\right)\left(g_{\mu \nu'}+2n_\mu n_{\nu'}\right)\right].\nn 
\end{align}
The $1/\epsilon$ pole has cancelled and the result is finite and unambiguous.

Finally, plugging in the expressions \eqref{conftensorcoords} for these bi-tensors in our conformally flat coordinate system \eqref{dsmetrice}, we get
\begin{align}
\begin{split} 
 \langle F_{\mu\nu}(x)F_{\mu'\nu'}(x')\rangle 
=&\    \frac{\Gamma(\Delta)}{\pi^{\Delta}} \frac{1}{|x-x'|^{2\Delta}} \left(I_{\mu\mu'}I_{\nu\nu'}- I_{\mu\nu'}I_{\nu\mu'}\right),
\end{split}
\end{align}
with $\Delta = D/2 = 2$ in $D = 4$ and $I_{\mu\nu'}(x,x')$ is the conformal structure \eqref{confstruct}.
This expression is exactly the conformally invariant two-point function of a rank $r=2$ antisymmetric, dimension $\Delta=2$ primary field in flat space.  By \eqref{deltatodwe}, the Weyl weight of such a field is $\Delta_W = \Delta - r = 0$, so this conforms with the general expression \eqref{weylrelatedcorree} and shows directly that the Maxwell theory is conformal on dS$_4$.

In flat space, it was shown in \cite{El-Showk:2011xbs} that in $D>4$ the scale but not conformally invariant Maxwell theory can be embedded into a larger non-unitary conformal theory.  This was possible because the correlators of $F_{\mu\nu}$, though not conformal in $D>4$, take the form of a descendent of a conformal primary of spin-1, and so by adding a (necessarily non-unitary) spin-1 primary to the theory, one obtains a non-unitary conformal theory with the Maxwell theory as a unitary but non-conformal subsector.  On (A)dS space, we find that this construction does not work.  For example, in $D=6$, taking a descendent of the conformal structure \eqref{confAcorr}, we obtain a correlator of the form
\be \langle \nabla_{[\mu}{A_{\nu]}}(x) \nabla_{[\mu'}{A_{\nu']}}(x') \rangle  \sim {1\over (Z-1)^3}\left[\left( g_{\mu\mu'}+{Z+5\over 2}n_\mu n_{\mu'}\right)\left( g_{\nu\nu'}+{Z+5\over 2}n_\nu n_{\nu'}\right)-(\mu\leftrightarrow \nu)\right].\ee
On the other hand, the massless vector correlator in $D=6$ gives
\be \langle F_{\mu\nu}(x) F_{\mu'\nu'}(x')\rangle  \sim {Z-3\over (Z-1)^3}\left[\left( g_{\mu\mu'}+{Z-7\over Z-3}n_\mu n_{\mu'}\right)\left( g_{\nu\nu'}+{Z-7\over Z-3}n_\nu n_{\nu'}\right)-(\mu\leftrightarrow \nu)\right].\ee
These have different tensor structures.  In the flat limit where $Z\rightarrow 1$, both reduce to eq. 3.2 of \cite{El-Showk:2011xbs}, but they differ away from flat space.

\section{Tensors\label{spin2section}}

A spin-2 field on dS$_D$ with mass $m$ is described by the  Fierz-Pauli Lagrangian \cite{Fierz:1939ix} extended to curved space (see section 5 of \cite{Hinterbichler:2011tt} for a review),
\bea
\begin{split}
{1\over \sqrt{-g}}{\cal L}_{{\rm FP},m^2}(h)=&\ -{1\over 2}\nabla_\rho h_{\mu\nu} \nabla^\rho h^{\mu\nu}+\nabla_\rho h_{\mu\nu} \nabla^\nu h^{\mu\rho}-\nabla_\mu h\nabla_\nu h^{\mu\nu}+\half \nabla_\mu h\nabla^\mu h \\
& +\left(D-1\right)H^2\left( h^{\mu\nu}h_{\mu\nu}-\half h^2\right)-\frac{1}{2}m^2(h_{\mu\nu}h^{\mu\nu}-h^2)\,.
\label{spin2lagegfpe} 
\end{split}
\eea

For generic $m^2$, this has no gauge symmetry and describes a massive spin-2 particle with ${D(D-1)\over 2}-1$ propagating degrees of freedom.
When $m^2=0$ we have the linear diffeomorphism gauge symmetry
\be \delta h_{\mu\nu}=\nabla_\mu \xi_\nu+\nabla_\nu\xi_\mu\, ,\label{linearggsyme}\ee
with vector gauge parameter $\xi^\mu(x)$, and the theory describes a massless spin-2 field with ${ D(D-3)\over 2} $ propagating degrees of freedom.  When $m^2=(D-2)H^2$ we have the double derivative gauge symmetry
\be \delta h_{\mu\nu}=\nabla_\mu \nabla_\nu\Lambda+H^2 g_{\mu\nu} \Lambda\,,\label{PMsymmetrye}\ee
with scalar gauge parameter $\Lambda(x)$, and the theory describes a partially massless (PM)  \cite{Deser:1983mm,Deser:2001pe} spin-2 field with ${D(D-1)\over 2}-2$ propagating degrees of freedom.

In what follows, we assume $D\geq4$, avoiding some potentially interesting degenerate cases in lower dimensions.

\subsection{Conformal symmetry\label{Confsymspin2}}
The conformal transformation \eqref{gconfrftraedsse} acting on $h_{\mu\nu}$ reads
\be 
\delta h_{\mu\nu}=-\left[ \xi^\rho \partial_\rho h_{\mu\nu} + h_{\mu\rho}\partial_\nu \xi^\rho + h_{\rho\nu}\partial_\mu \xi^\rho +{\Delta-2\over D} \left( \partial\cdot \xi -{D\over 2} { H^2 \xi\cdot x\over 1+{H^2\over 4} x^2}\right)h_{\mu\nu} \right].\label{hconftrans}
\ee
The Fierz-Pauli theory \eqref{spin2lagegfpe} is not invariant under this transformation for any value of $m^2$ or $\Delta$, so it is not conformally invariant at the level of the action when $h_{\mu\nu}$ transforms in the usual way.  This includes the massless graviton.  It also includes the partially massless graviton, consistent with the claims of \cite{Barnich:2015tma} (note, however, that the free PM graviton in $D=4$ is invariant under a non-standard version of conformal symmetry \cite{Deser:2004ji,Letsios:2023tuc}, whose nature has yet to be fully elucidated).  As we will see later, the massless graviton is in fact conformal at the level of correlation functions of gauge invariant local operators only when $D=4$, as happens in the flat space case \cite{Farnsworth:2021zgj}.  For the partially massless graviton, there is no conformal invariance in any $D$ even at the level of correlators of gauge invariant local operators, as argued in \cite{Barnich:2015tma}.

As in the vector case, there is a conformal theory with action invariant under \eqref{hconftrans} if we allow for a non-canonical kinetic term.  Considering the general action for a symmetric tensor $h_{\mu\nu}$ on dS$_D$ with up to two derivatives, which we can write in the form,
\bea {1\over \sqrt{-g}}{\cal L}_{h}= {1\over \sqrt{-g}}{\cal L}_{{\rm FP},m^2}(h) +\xi_1  \nabla^\nu h_{\mu\nu} \nabla_\rho  h^{\mu\rho}+\xi_2 \nabla_\mu h\nabla_\nu h^{\mu\nu}+\xi_3 \nabla_\mu h\nabla^\mu h + a H^2 h^2\,. 
\label{spin2lagege} 
\eea
The is the Fierz-Pauli Lagrangian \eqref{spin2lagegfpe} along with terms parametrizing the possible deviations from the Fierz-Pauli Lagrangian, with parameters $\xi_1,\xi_2,\xi_3,a$.

Demanding conformal invariance under \eqref{hconftrans}, we find a one parameter family of solutions\footnote{In writing \eqref{spin2lagege} we have implicitly assumed that the leading structure $(\nabla_\rho h_{\mu\nu})^2$ is non-vanishing.  Relaxing this assumption does not yield any other non-trivial solutions.},
\begin{align} 
\begin{split}
& \xi_1=-{(D-2)\over (D+2)},\ \ \xi_2={(D+4)(D-2)\over D(D+2)},\ \ \xi_3= - \frac{(D-2)(D^2+3D+4)}{2D^2(D+2)} -{1\over 2}\lambda ,\\
& {m^2\over H^2}=\frac{D (D-2)  }{4},\ \ \ a=-\frac{(D-1)(D-2) (D-4)}{8D}-{D(D-2)\over 8}\lambda\,, \ \ \ \Delta={D\over 2}-1\,,\label{confospin2sole} 
\end{split}
\end{align}
where $\lambda$ is a free parameter.\footnote{\label{lambdavaluefootnote}For $\lambda = - \frac{(D-2)(D-4)}{2D^2}$, the values of $\xi_1,\ \xi_2$ and $\xi_3$ are such that these three terms can be arranged into the form
\begin{align}
-\frac{(D-2)}{(D+2)}\left( \nabla_\nu h^{\mu\nu} - \frac{(D+4)}{2D} \nabla^\mu h\right)^2,
\end{align}
which is precisely the gauge fixing term needed in the flat space case for the massless action to be conformal in $D \neq 4$ \cite{Farnsworth:2021zgj}.}

We can make clear the reason for the free parameter $\lambda$ by decomposing the field into a traceless part and a trace,
\be h_{\mu\nu}=\bar h_{\mu\nu}+{1\over D}\varphi g_{\mu\nu},\ \ \  \bar h^\mu_{\ \mu}=0, \ \ \varphi=h\, ,\ee
after which the conformal action decouples into two parts,
\be {\cal L}_h={\cal L}_\varphi+{\cal L}_{\bar h}\, ,\ee
with
\begin{align}
  {1\over \sqrt{-g}}{\cal L}_{\bar h}&=   -{1\over 2}\nabla_\rho \bar h_{\mu\nu} \nabla^\rho \bar h^{\mu\nu}+{4\over D+2} \nabla_\rho\bar h_{\mu\nu} \nabla^\nu \bar h^{\mu\rho} -\frac{D^3-28 D+16}{8 (D+2)} H^2\bar  h^{\mu\nu}\bar h_{\mu\nu}  \,, \label{tildehlage} \\
 {1\over \sqrt{-g}}{\cal L}_{\varphi}&=\lambda \left[-\frac{1}{2}(\nabla\varphi)^2-{D(D-2)\over 8}H^2 \varphi^2 \right].
\end{align}
These are separately conformally invariant: ${\cal L}_\varphi$ is $\lambda$ times the conformal scalar Lagrangian (\eqref{scalaracele} with the mass value in \eqref{scalarconfvaluee}) and ${\cal L}_{\bar h}$ is the unique conformal second-order Lagrangian for a symmetric traceless tensor field.  Only the sign of $\lambda$ is meaningful because there is the field redefinition $h_{\mu\nu}\rightarrow h_{\mu\nu}+c\,h g_{\mu\nu}$ with constant $c$ which preserves the ansatz \eqref{spin2lagege} and can be used to scale the value of $\lambda$ by a positive number proportional to $ c^2$.
A value $\lambda\not=0$ indicates that the scalar mode is present, with the sign of its kinetic term determined by the sign of $\lambda$.  When $\lambda=0$, the scalar is absent and the trace disappears from the Lagrangian, indicating the presence of the linear Weyl symmetry 
\be \delta h_{\mu\nu}=\Omega_1 g_{\mu\nu}\, ,\label{linearwsyee}\ee
in ${\cal L}_h$, with arbitrary scalar gauge parameter $\Omega_1(x)$.  

In $D>4$, ${\cal L}_{\bar h}$ in \eqref{tildehlage} is a dS version of the $\kappa=1$, $s=2$ ``special'' conformal field in the classification of \cite{Metsaev:2016oic}, also discussed in \cite{Deser:1983mm,Iorio:1996ad,Erdmenger:1997wy,Gover:2008pt,Gover:2008sw,Faci:2012yg,BenAchour:2013zwh,Queva:2015vaa,Aros:2022ecb}.  It comes from the following Weyl invariant Lagrangian \cite{Leonovich:150772}, where $\bar h_{\mu\nu}$ transforms with Weyl weight $\Delta_W={D\over 2}-3$,
\be 
\begin{split}  {1\over \sqrt{-g}}{\cal L}_{{\rm Weyl},\bar h}=&\  -{1\over 2}\nabla_\lambda \bar h_{\mu\nu} \nabla^\lambda \bar h^{\mu\nu}+{4\over D+2} \nabla_\lambda\bar h_{\mu\nu} \nabla^\nu \bar h^{\mu\lambda} \\
& +{2\over D+2}\left[ R_{\mu\nu} -\frac{D^2-12}{16 (D-1)}R g_{\mu\nu}\right] \bar h^{\mu\rho}\bar h^{\nu}_{\ \rho} +b\, W_{\mu\rho\nu\sigma}\bar h^{\mu\nu}\bar h^{\rho\sigma} , 
\end{split} \label{rtensoracelegke2ewe}\ee 
which reduces to \eqref{tildehlage} when restricted to dS space.  (Here $b$ is an arbitrary constant, it comes with a term that is separately Weyl invariant, involving the Weyl tensor $W_{\mu\rho\nu\sigma}$. This term vanishes on conformally flat spaces such as dS.)

The equations of motion coming from \eqref{tildehlage} are
\be  \left( \nabla^2 -\frac{(D-2)^2 (D+4) }{4 (D+2)} H^2 \right) \bar h_{\mu\nu} -{8\over D+2} \nabla_{(\mu} \nabla^\rho \bar h_{\nu)_T\, \rho}=0.\ee
Taking a divergence gives
\be {D-2\over D+2}\left[ \left( \nabla^2 -{D^2-2D+4 \over 4}H^2\right) \nabla^\nu \bar h_{\mu\nu} -{4\over D} \nabla_\mu \nabla_\nu\nabla_\rho \bar h^{\nu\rho}\right]=0.\ee
This is the conformal vector equation \eqref{confvecteqe} for the quantity $ \nabla^\nu \bar h_{\mu\nu}$.  It follows that the double divergence $\nabla^\mu \nabla^\nu \bar h_{\mu\nu}$ satisfies the scalar wave equation with the conformal mass in \eqref{scalarconfvaluee}.  The analogous divergences on the equations for a pure massive spin-2 would have yielded constraints, so we expect to find an extra propagating vector and scalar mode in the case here.  

Returning to our original action \eqref{spin2lagege} with \eqref{confospin2sole}, we can explicitly isolate these extra scalar and vector modes as follows.  First introduce an auxiliary scalar $\phi$ and auxiliary vector $A_\mu$ to the Fierz-Pauli theory \eqref{spin2lagegfpe} with $m^2 = D(D-2)/4\, H^2$,
\begin{align}
\begin{split}
  {1\over \sqrt{-g}}{\cal L}=&\  {1\over \sqrt{-g}}{\cal L}_{{\rm FP},{D(D-2)\over 4}H^2}(h)+{D(D-1)(D-2)(D-4)\over 16}H^2\left( -h\phi+{D(D-1)\over 4} \phi^2\right) \\ 
&+{D(D-2)\over 2}H^2 \left(  - A_\mu \nabla_\nu h^{\mu\nu} +{D+4\over 2D} A_\mu\nabla^\mu h +{D(D+2)\over 8}H^2A^2 \right). \label{phihaacrtioe}
\end{split}
\end{align}
The $\phi$ and $A_\mu$ equations of motion give
\be \phi={2\over D(D-1)}h, \ \ A_\mu={4\over D(D+2)H^2} \left(\nabla^\nu h_{\mu\nu}- {D+4\over 2 D}\nabla_\mu h\right),\ee
which upon plugging back into \eqref{phihaacrtioe} gives the solution \eqref{confospin2sole} with $\lambda=- \frac{(D-2)(D-4)}{2D^2}$ (see footnote \ref{lambdavaluefootnote}).
We can now diagonalize \eqref{phihaacrtioe} by making the field redefinitions
\begin{align}
\begin{split}
 h_{\mu\nu}&\rightarrow \tilde{h}_{\mu\nu}+ \nabla_{(\mu}A_{\nu)}  +\frac{1}{H^2} \left[  \nabla_\mu\nabla_\nu\tilde{\phi}   +{D\over 4}H^2 \tilde{\phi} \, g_{\mu\nu} \right],\\
\phi &\rightarrow \tilde{\phi}+\frac{2}{D(D-1)} \nabla\cdot A\,,
\end{split}
\label{spin2fieldredef}
 \end{align}
after which we have
\begin{align}
  {1\over \sqrt{-g}}{\cal L}=&\  {1\over \sqrt{-g}}{\cal L}_{{\rm FP},{D(D-2)\over 4}H^2}(\tilde{h})-{D(D-2)\over 8}H^2\left[-{1\over 4}F_{\mu\nu}^2-{(D-4)\over 2D} (\nabla\cdot A)^2-{(D-4)(D-2)\over 8}H^2A^2\right] \nn\\
& -\frac{D (D-1)(D-2) (D-4)}{16}  \left[ -\frac{1}{2} (\nabla\tilde{\phi})^2- {D(D-2)\over 8}H^2\tilde{\phi}^2\right]. \label{phihaacrtioe2}
\end{align}

The Lagrangian for $A_\mu$ appearing here is proportional to the conformal vector theory \eqref{vescalaracelegke2}, which can itself be diagonalized into a massive vector and a conformal scalar $\phi_A$ as we did in section \ref{vectorsection}.  The Lagrangian for $\tilde{\phi}$ is another conformal scalar.  In total we have a massive tensor, massive vector and two conformal scalars.  One of the conformal scalars is the trace mode present when $\lambda\neq 0$, so the case $\lambda=0$, or equivalently the traceless Lagrangian \eqref{tildehlage}, therefore contains a massive tensor, massive vector, and a single massive scalar, with masses,
\be m^2_{\tilde{h}}=  {D(D-2)\over 4}H^2,\  \ \ m_{A}^2={(D-4)(D-2)\over 4}H^2 , \ \ m^2_{\phi_A}=  {D(D-2)\over 4}H^2.\label{tesvtmassese}\ee
Note that these fields all have the same dual CFT dimension $\delta_\pm={d\pm 1\over 2}$, where $d \equiv D-1$, so they can be considered as generalizations of the conformal mass of the scalar.
As in the vector case in section \ref{vectorsection}, because of the diagonalization \eqref{spin2fieldredef}, the conformal symmetry mixes the tensor, vector and scalar modes, but still acts linearly, so these fields form an irreducible representation of the conformal algebra which break up into three representations when restricted to the isometry algebra.

On dS, the mass for $\tilde h_{\mu\nu}$ is above the Higuchi bound \cite{Higuchi:1986py} and the vector $A_\mu$ and scalar $\phi_A$ masses are positive, but the signs of the kinetic terms for these three fields alternate.  On AdS, the relative signs of these three fields are all the same, but all the fields have negative masses squared, and so the theory is non-unitary.  Either way, the full theory is not unitary on dS or AdS.

\subsubsection{$D = 4$}

An additional interesting symmetry appears in this generalized conformal theory when $D=4$.  In this case the conformal mass for the tensor in \eqref{tesvtmassese} coincides with the partially massless value $m^2 = (D-2)H^2$, and the vector becomes massless. The action \eqref{spin2lagege} with parameters \eqref{confospin2sole} and $\lambda = 0$  is invariant under the double derivative gauge transformation
\be \delta h_{\mu\nu}=\nabla_{\mu}\nabla_{\nu}\Omega_2\, ,\label{PMD4wsymee} \ee
for arbitrary scalar gauge parameter $\Omega_2(x)$, in addition to the Weyl symmetry \eqref{linearwsyee}.
{The Lagrangian in this case can be written as\footnote{ The standard PM action (the Fierz-Pauli theory \eqref{spin2lagegfpe} with $m^2=(D-2)H^2$) can be written as
\be {1\over \sqrt{-g}}{\cal L}_{\rm PM}=-{1\over 4}\left(F_{\mu\nu\rho}F^{\mu\nu\rho}-2 F_{\mu}F^{\mu}\right),\label{PMintffe}\ee
where $F_{\mu\nu\rho}\equiv\nabla_\mu h_{\nu\rho} -\nabla_\nu h_{\mu\rho}$ is the field strength invariant under the PM symmetry \eqref{PMsymmetrye} and $F_{\mu}\equiv F_{\mu\nu}^{\ \ \nu}$ is its trace \cite{Deser:2006zx}.  The precise relative coefficient between the two structures in \eqref{PMintffe}   ensures that the divergence of the equations of motion yields the constraint $F_{\mu}=0$, which removes unwanted extra degrees of freedom and allows only the PM spin-2 degrees of freedom to propagate.  The $D$ dimensional version of the Lagrangian \eqref{lincogee}, using the $D$ dimensional expression $C_{\mu\nu\rho}=F_{\mu\nu\rho}-{2\over D-1}g_{\rho[\mu}F_{\nu]}$, instead has the structure
\be {1\over \sqrt{-g}}{\cal L}_{}=-{1\over 4}C_{\mu\nu\rho}C^{\mu\nu\rho}=-{1\over 4}\left(F_{\mu\nu\rho}F^{\mu\nu\rho}+{2\over D-1} F_{\mu}F^{\mu}\right).\label{PMintffe2}\ee
This no longer has a divergence constraint, but instead has the Weyl symmetry \eqref{linearwsyee} (as was noted in \cite{Garcia-Saenz:2014cwa}), and as a result it propagates an extra massless vector mode.   And unlike the pure PM theory, it is conformally invariant in $D=4$.
}

\be {1\over \sqrt{-g}}{\cal L}=-{1\over 4}C_{\mu\nu\rho}C^{\mu\nu\rho}\, ,\label{lincogee}\ee
where 
\be C_{\mu\nu\rho}\equiv\nabla_\mu h_{\nu\rho} -\nabla_\nu h_{\mu\rho}-{2\over 3}g_{\rho[\mu}\left(\nabla^\sigma h_{\nu]\sigma}-\nabla_{\nu]} h\right).\label{Cdefinitione}\ee
The tensor $C_{\mu\nu\rho}$ is invariant under \eqref{PMD4wsymee} and \eqref{linearwsyee}.  It is fully traceless, antisymmetric in its first two indices, and vanishes if anti-symmetrized over all its indices, i.e. it has the symmetries of a traceless hook tableau $\raisebox{5pt}{\resizebox{.5cm}{!}{\gyoung(\mu\rho ,\nu )}}$,
\be C_{\mu\nu\rho}=-C_{\nu\mu\rho} \,,\ \ \ C_{[\mu\nu\rho]}=0\, , \ \ \ C^{\ \ \  \nu}_{\mu\nu}=0\,. \ee
This is the basic gauge-invariant local operator in the theory, which we will compute correlators of in section \ref{tensorcorrelatorsec}.
We can think of \eqref{lincogee} as analogous to linearized conformal gravity, with \eqref{Cdefinitione} analogous to the linearized Weyl tensor. }
 
Going to the traceless Lagrangian ${\cal L}_{\bar h}$ \eqref{tildehlage} by using the Weyl symmetry to set the trace to zero, we are left with the gauge symmetry
\be \delta \bar h_{\mu\nu}=\nabla_{(\mu}\nabla_{\nu)_T}\Omega_2\, ,\label{PMD4wsymtee} \ee
inherited from \eqref{PMD4wsymee}.  This case is the (A)dS version of the $s= 2$, $\kappa=1$ ``type II partial-short'' conformal field in the classification of \cite{Metsaev:2016oic}. It appears in \cite{Drew:1980yk}, 
and is studied further, along with its curved space extensions, in \cite{Barut:1982nj,Anselmi:1999bu,Queva:2015vaa}.

To diagonalize this theory, we can work with the Weyl invariant theory because the value $\lambda=- \frac{(D-2)(D-4)}{2D^2}$ that we needed in order to facilitate the diagonalization coincides with the value $\lambda= 0$ where we also have invariance under Weyl transformations \eqref{linearwsyee}.  Furthermore, we do not need to introduce the auxiliary scalar since the scalar terms in \eqref{phihaacrtioe} all vanish in $D=4$.
 We introduce the vector $A_\mu$ as in \eqref{phihaacrtioe} with $D=4$, then diagonalize via the field redefinition
\begin{align}
 h_{\mu\nu}&\rightarrow \tilde{h}_{\mu\nu}+ \nabla_{(\mu}A_{\nu)},\label{PMfieldredefe}
 \end{align}
giving the Lagrangian 
\begin{align}
  {1\over \sqrt{-g}}{\cal L}=&\  {1\over \sqrt{-g}}{\cal L}_{{\rm FP},2H^2}(\tilde{h})-H^2\left[-{1\over 4}F_{\mu\nu}^2\right]. \label{confd4lage}
\end{align}
The spin-2 field $\tilde{h}_{\mu\nu}$ now has the partially massless value $m^2=2H^2$, the vector $A_\mu$ is massless, and there is no scalar.  Thus in $D=4$ there is a conformal model consisting of a PM spin-2 field and a massless spin-1 field which transform into each other under conformal transformations.   

In terms of the diagonalized fields, the two gauge symmetries  \eqref{linearwsyee}, \eqref{PMD4wsymee} become
\be \delta \tilde{h}_{\mu\nu}=\nabla_\mu\nabla_\nu\Lambda_1+H^2 g_{\mu\nu}\Lambda_1,\ \ \ \ \delta A_\mu=\nabla_\mu\Lambda_2\, ,
\label{PMgaugesym}\ee
where
\be \Lambda_1={1\over H^2}\Omega_1\, , \ \ \  \Lambda_2=-{1\over 2 H^2} \Omega_1+{1\over 2} \Omega_2 \, .\ee
These are precisely the PM gauge transformations \eqref{PMsymmetrye} and $U(1)$ gauge transformations \eqref{emgaugesyme} of the PM graviton and massless vector, respectively.

On dS$_4$, where the PM field is unitary, there is a wrong relative sign between the PM tensor  Lagrangian and the massless vector Lagrangian.  On AdS$_4$, the relative sign is correct, but the PM field itself is not unitary.  Either way, the model is not unitary (non-unitarity in the flat space case was noted in \cite{Fang:1982ks}).  Note that this is the same as the structure of conformal gravity: conformal gravity expanded around (A)dS propagates a partially massless graviton and a massless graviton, and its unitarity properties are the same as was just described with the photon replaced by the massless graviton \cite{Maldacena:2011mk,Deser:2012qg,Deser:2012euu,Joung:2014aba}.

From the arguments of \cite{Barnich:2015tma}, a necessary condition for gauge theories to be conformal is that the reducibility parameters,  i.e. the global part of the gauge symmetries, should form representations of the conformal algebra $\frak{so}(2,4)$.  The reducibility parameters always form multiplets of the dS isometry algebra $\frak{so}(1,4)$, but it is a non-trivial requirement that they combine into representations of the larger conformal algebra.  In the case at hand here, we can see that this condition is met: the reducibility parameters of the partially massless tensor field form a vector of the dS isometry algebra $\frak{so}(1,4)$ \cite{Hinterbichler:2015nua}, whereas the reducibility parameter of the massless vector field is a scalar.   Together, these join into a vector of the conformal algebra $\frak{so}(2,4)$.

\subsection{Correlator\label{tensorcorrelatorsec}}

We will consider first the correlator of the Fierz-Pauli theory \eqref{spin2lagegfpe}, which we will then use to compute the correlator of the conformal theories and the massless graviton.  The two-point function for a massive graviton in dS$_D$ has been calculated in \cite{DHoker:1999bve,Buchbinder:1999be,Buchbinder:1999ar, Bena:1999be,Naqvi:1999va,Costa:2014kfa}.
The two-point function of a symmetric tensor $h_{\mu\nu}$ on Euclidean dS can be decomposed into a linear combination of symmetric bi-tensors as follows:
\begin{align}
\begin{split}
 \langle h_{\mu\nu}(x) h_{\mu'\nu'}(x')\rangle =&\ f_1(Z) \, g_{\mu \nu} g_{\mu '\nu'}\\
&+f_2(Z) \left( g_{\mu \mu '} g_{\nu\nu'} + g_{\mu \nu'} g_{\nu\mu '}\right)\\
&+f_3(Z)\, n_\mu  n_\nu n_{\mu '} n_{\nu'}\\
&+f_4(Z)\left( g_{\mu \nu} n_{\mu '}n_{\nu'} + g_{\mu '\nu'} n_\mu  n_\nu\right)\\
&+f_5(Z)\left( g_{\mu \mu '} n_\nu n_{\nu'} + g_{\mu \nu'} n_\nu n_{\mu '} + g_{\nu \nu'} n_\mu  n_{\mu '} + g_{\nu\mu '} n_\mu  n_{\nu'}\right).
\end{split} \label{massgravtwopt2}
\end{align}
{The equations of motion following from \eqref{spin2lagegfpe} are equivalent to 
\be \left(\nabla^2 -2H^2-m^2\right)h_{\mu\nu}=0\, ,\ \ \ \nabla^\nu h_{\mu\nu}=0\, ,\ \ h^\mu_{\ \mu}=0\, .\ee
Imposing these on \eqref{massgravtwopt2} at separated points fixes} 
\begin{align} 
f_1(Z) &=C_{D,m^2} \left[ G''(Z)-  {m^2\over H^2} {2\over (D-2)}  Z G'(Z)- {m^2\over H^2} \left(1+{m^2\over (D-2) H^2}\right) G(Z)\right],\nn \\
f_2(Z) &=  C_{D,m^2} \left[ G''(Z)+{m^2\over H^2}{(D-1)\over (D-2)}  Z G'(Z)+ {m^2\over H^2} {(D-1)\over 2} \left(1+{m^2\over (D-2)H^2}\right) G(Z) \right], \nn\\
f_3(Z) &= C_{D,m^2} \bigg[\left( 4 (D+1) Z+\left(D^2+2 D+4\right)\right) G''(Z) \\ & \quad\quad\quad\quad\quad +2 \left({m^2\over H^2} Z  + (D+1) \left(D+{m^2\over H^2}\right)\right) G'(Z) + {m^2\over H^2} (D-2) \left(1+{m^2\over (D-2)H^2}\right) G(Z) \bigg], \nn\\
f_4(Z) &= C_{D,m^2} \left[-(D+2) G''(Z)+ {m^2\over H^2}{2\over (D-2)}  Z G'(Z)+ {m^2\over H^2} \left(1+{m^2\over (D-2)H^2}\right) G(Z)\right],\nn\\
f_5(Z)&=C_{D,m^2} \bigg[( (D+1)  Z+1) G''(Z)+\left({m^2\over H^2}{(D-1)\over (D-2)}  Z+ {(D+1)\over 2} \left(D+{m^2\over H^2}\right)\right) G'(Z) \nn\\ & \quad\quad\quad\quad\quad +{m^2\over H^2}{(D-1)\over 2}  \left(1+{m^2\over (D-2)H^2}\right) G(Z)\bigg],\nn
\end{align}
where 
\begin{align}
C_{D,m^2} \equiv \frac{(D-2)}{(D-1)}\frac{1}{\frac{m^2}{H^2} (\frac{m^2}{H^2}-D+2)}\,,
\end{align}
and the function $G(Z)$ obeys the equation
\begin{align}
 (Z^2-1) G''(Z) +DZ G'(Z)+\frac{m^2}{H^2} G(Z) &= 0\,. \label{Gequamgre}
\end{align}

The equation \eqref{Gequamgre} is exactly the equation \eqref{scZee} in the scalar case, so the solution in Euclidean dS with appropriate boundary conditions is the same as \eqref{genscalarproze},
\be G(Z)={H^{D-2}\Gamma(\delta_+)\Gamma(\delta_-)\over 2^D \pi^{D/2} \Gamma\left(D/2\right)}\, {_{2}F_1} \left( \delta_-,\delta_+;{D\over 2};{Z+1\over 2}\right),\ee
with
\be \textnormal{spin-2:}\quad{} \delta_\pm \equiv {d\over 2}\pm\sqrt{{d^2\over 4}-{m^2\over H^2}}\, ,\ \ \ d\equiv D-1\,.\ee
Here $\delta_\pm$ are the dual CFT conformal dimensions for a massive spin-2 field with the mass $m$.

As a function of the mass, the correlator becomes singular at the massless value $m^2=0$, where the theory becomes gauge invariant under \eqref{linearggsyme}, and the PM value $m^2=(D-2)H^2$, where the theory becomes gauge invariant under \eqref{PMsymmetrye}.  These singularities in the propagator will cancel in correlators of gauge invariant observables.

{The correlator also becomes singular at the mass values
\be m^2  = -(k+2)(k+D+1)H^2\, ,\ \ \ k=0,1,2,\ldots\,  , \ee
for which $\delta_+=d+k+2$, $\delta_-=-k-2$.  {These singularities come from the poles of $\Gamma(\delta_-+2)$, which appear in $G''(Z)$.} These are the values at which the massive graviton acquires enhanced shift symmetries \cite{Bonifacio:2018zex}.  At these values, the shift symmetries should be gauged and correlators of the shift invariant operators will be finite. 
}

\subsubsection{Conformal theory in $D > 4$}

In $D>4$, the linearized spin-2 action around dS$_D$ with non-canonical kinetic terms \eqref{spin2lagege} is conformally invariant with the choice \eqref{confospin2sole}.  Here we will calculate its two-point function and show directly that it is conformally invariant. As shown in section \ref{Confsymspin2}, this action is equivalent via the field redefinitions \eqref{spin2fieldredef} to \eqref{phihaacrtioe2} with the conformal masses \eqref{tesvtmassese}, which is a massive spin-2 Fierz-Pauli action, a massive scalar, and a non-canonical conformal massive vector of the type considered in section \ref{vectorsection}.  To calculate the two-point function of \eqref{spin2lagege}, \eqref{confospin2sole}, we only need the two-point function for the massive spin-2 field, and our previous results for the two-point function of the conformal vector and scalar.  With these we can reconstruct the two-point function of the original conformal tensor using the field redefinitions \eqref{spin2fieldredef}.

For the particular conformal value of the mass $m^2_{\tilde{h}}=  {D(D-2)\over 4}H^2$, the massive graviton two-point function from above simplifies dramatically, giving
\begin{align}
\langle \tilde h_{\mu\nu}(x)\tilde h_{\mu'\nu'}(x')\rangle 
=&\ \frac{ H^{D-2}\Gamma(D/2-1)}{2(D-1)(D-4)(2\pi)^{D/2} } \frac{1}{(1-Z)^{D/2+1}}\label{gravtilde2pt} \\
&\times \bigg[\left(4Z-D(1-Z)^2\right)g_{\mu \nu} g_{\mu '\nu'}\nn\\
&\ \ \ +\frac{1}{2} \left(4(D+1)-4(D-1)Z+D(D-1)(1-Z)^2\right) \left( g_{\mu \mu '} g_{\nu\nu'} + g_{\mu \nu'} g_{\nu\mu '}\right)\nn\\
&\ \ \ +\left(8(D+1)(D+2) -4(D^2-4)Z+D(D -2)(1-Z)^2\right)n_\mu  n_\nu n_{\mu '} n_{\nu'}\nn\\
&\ \ \ -\left(2(D+2)+D(1-Z) \right)(Z+1)\left( g_{\mu \nu} n_{\mu '}n_{\nu'} + g_{\mu '\nu'} n_\mu  n_\nu\right)\nn\\
&\ \ \ +\frac{1}{2}\left(8(D+2)-2(D^2+D-4)Z+ D(D-1)(1-Z)^2\right)\nn\\
&\ \ \ \ \ \times \left( g_{\mu \mu '} n_\nu n_{\nu'} + g_{\mu \nu'} n_\nu n_{\mu '} + g_{\nu \nu'} n_\mu  n_{\mu '} + g_{\nu\mu '} n_\mu  n_{\nu'}\right)\bigg].\nn
\end{align}
To find the full two-point function for $h_{\mu\nu}$, we use the first equation of \eqref{spin2fieldredef} used to diagonalize the action,
\begin{align}
\begin{split}
 h_{\mu\nu}&= \tilde{h}_{\mu\nu}+ \nabla_{(\mu}A_{\nu)}  +\frac{1}{H^2} \left[  \nabla_\mu\nabla_\nu\tilde{\phi}   +{D\over 4}H^2 \tilde{\phi} g_{\mu\nu} \right].
\end{split}
\label{redefundo}
 \end{align}
 
 Since the actions for the scalar $\tilde{\phi}$ and vector $A_\mu$ are exactly the conformal actions we saw in the previous sections, we can use the two-point functions for these fields we computed there, modulo the appropriate normalizations.   The scalar correlator from \eqref{confscorre} gives
 \begin{align}
 \begin{split}
 \langle \tilde{\phi}(x) \tilde{\phi} (x')\rangle= -\frac{16}{D(D-4)(D-2)(D-1)} {H^{D-2}\Gamma(D/2)\over (D-2) (2\pi)^{D/2} }\frac{1}{(1-Z)^{D/2-1}}\,,
 \end{split}
 \end{align}
and the vector correlator from \eqref{confAcorr} gives
 \begin{align}
 \begin{split}
&\ \langle A_\mu(x) A_{\nu'}(x')\rangle =-\frac{8}{D(D-2)H^2} \frac{ H^{D-2}\Gamma(D/2)}{(D-4) (2 \pi)^{D/2}}\frac{1}{(1-Z)^{D/2-1}} \left(g_{\mu\nu'}+  2 n_\mu n_{\nu'}\right).
\end{split}
\end{align}

 We can now assemble the correlation function of ${h}_{\mu\nu}$ from \eqref{redefundo}, 
  \begin{align}
 \langle h_{\mu\nu}(x) h_{\mu'\nu'}(x')\rangle  =&\ \langle \tilde h_{\mu\nu}(x) \tilde h_{\mu'\nu'}(x')\rangle  \\
 &+\frac{1}{4}\left(\nabla_\mu \nabla_{\mu'}\langle A_\nu(x) A_{\nu'}(x')\rangle +\nabla_\mu \nabla_{\nu'}\langle A_\nu(x) A_{\mu'}(x')\rangle +\mu \leftrightarrow \nu \right) \nn\\
 &+ \frac{1}{H^4}  \nabla_\mu\nabla_\nu\nabla_{\mu'}\nabla_{\nu'} \langle \tilde \phi (x) \tilde \phi (x')\rangle + \frac{D}{4}\frac{1}{H^2}  \left( g_{\mu'\nu'} \nabla_\mu\nabla_\nu+g_{\mu\nu} \nabla_{\mu'}\nabla_{\nu'}  \right)\langle \tilde \phi (x) \tilde \phi (x')\rangle\nn \\
 &+\frac{D^2}{16} g_{\mu\nu}g_{\mu'\nu'} \langle \tilde \phi (x) \tilde \phi (x')\rangle \,.\nn
 \end{align}
 The derivatives can be computed using the expressions in Appendix \ref{appendixbi-tensor} and we get:
\begin{align}
& \langle h_{\mu\nu}(x) h_{\mu'\nu'}(x')\rangle =  \frac{D}{(D-2)(D-4)}\frac{H^{D-2} \Gamma(D/2)}{(2 \pi)^{D/2} }\frac{1}{(1-Z)^{D/2-1}}\\
&\times \bigg[\frac{1}{2} \left(\left( g_{\mu \mu '}+2n_\mu n_{\mu'}\right)\left( g_{\nu\nu'} +2n_\nu n_{\nu'}\right)+\left( g_{\mu \nu'} +2n_\mu n_{\nu'}\right) \left( g_{\nu\mu '}+2 n_\nu n_{\mu'}\right) \right)-\frac{1}{(D-2)}g_{\mu \nu} g_{\mu '\nu'}\bigg]. \nonumber
\end{align}
Separating this into the traceless part $\bar{h}_{\mu\nu}$ and trace $\varphi$ using $h_{\mu\nu} = \bar{h}_{\mu\nu} +\frac{1}{D}g_{\mu\nu} \varphi$, we get for the traceless two-point function
\begin{align}
& \langle \bar h_{\mu\nu}(x) \bar h_{\mu'\nu'}(x')\rangle  =  \frac{D}{(D-2)(D-4)}\frac{H^{D-2} \Gamma(D/2)}{(2 \pi)^{D/2} }\frac{1}{(1-Z)^{D/2-1}}\label{hhconf} \\
&\times \bigg[\frac{1}{2} \left(\left( g_{\mu \mu '}+2n_\mu n_{\mu'}\right)\left( g_{\nu\nu'} +2n_\nu n_{\nu'}\right)+\left( g_{\mu \nu'} +2n_\mu n_{\nu'}\right) \left( g_{\nu\mu '}+2 n_\nu n_{\mu'}\right) \right)-\frac{1}{D}g_{\mu \nu} g_{\mu '\nu'}\bigg],\nonumber
\end{align}
and for the trace
 \begin{align}
 \langle \varphi(x) \varphi(x')\rangle =&\ -\frac{2D^2}{(D-2)^2(D-4)}\frac{H^{D-2} \Gamma(D/2)}{(2 \pi)^{D/2} }\frac{1}{(1-Z)^{D/2-1}} \,,
 \end{align}
 with the cross-correlator vanishing,
  \begin{align}
 \langle \bar{h}_{\mu\nu}(x) \varphi(x')\rangle =&\  0\, .
 \end{align}
 Finally, writing this in the conformally flat coordinates \eqref{dsmetrice}, we have
 \begin{align}
\langle \bar{h}_{\mu\nu}(x) \bar{h}_{\mu'\nu'}(x')\rangle =&\ \Omega(x)^{-\Delta+2} \Omega(x')^{-\Delta+2} \bigg[\frac{D\Gamma(\Delta)}{4(D-4)(\pi)^{D/2} }\left(\frac{1}{|x-x'|^{2\Delta}}\right)\mathcal{I}_{\mu\nu;\mu'\nu'}\bigg], \label{trtensorxeceae}
\end{align}
with 
\begin{align}
\Omega(x) = \frac{1}{\left(1+\frac{H^2}{4} x^2\right)},\quad{} \Delta = \frac{D}{2} - 1\,, 
\label{omegaDeltagrav}
\end{align}
and
\begin{align}
\mathcal{I}_{\mu\nu;\mu'\nu'} = \frac{1}{2} \left(I_{\mu\mu'} I_{\nu\nu'} +I_{\mu\nu'} I_{\nu\mu'} \right)-\frac{1}{D}g_{\mu \nu} g_{\mu '\nu'}\,
\end{align}
where $I_{\mu\mu'}$ is defined in \eqref{confstruct}. The trace two-point function in these coordinates becomes
\begin{align}
 \langle \varphi(x) \varphi(x')\rangle =&\ \Omega(x)^{-\Delta} \Omega(x')^{-\Delta} \left[\frac{ \Gamma(\Delta)}{2(D-2)(D-4)(\pi)^{D/2} }\left(\frac{1}{|x-x'|^{2\Delta}}\right)\right] \label{finalcorrdn4ee}
 \end{align}
 with $\Omega(x)$ and $\Delta$ as defined in \eqref{omegaDeltagrav}. The term in square brackets in \eqref{trtensorxeceae} has the structure of a two-point function of a traceless spin-2 primary with dimension $\Delta$ in flat space.  It gets multiplied by Weyl factors with Weyl weight $\Delta_W = \Delta - 2$ for the rank-2 field, confirming \eqref{weylrelatedcorree} and \eqref{deltatodwe} and demonstrating that the traceless non-canonical tensor theory is conformal on dS$_D$.  The terms in square brackets in \eqref{finalcorrdn4ee} has the  structure of a two-point function of a scalar primary with dimension $\Delta$ in flat space, and it gets multiplied by Weyl factors with Weyl weight $\Delta_W = \Delta $ appropriate for a scalar.

Note that both expressions \eqref{trtensorxeceae} and \eqref{finalcorrdn4ee} are singular in the $D \rightarrow 4 $ limit, a sign of the gauge invariance that develops in this case.  Below we will see that this singularity cancels when we compute the appropriate gauge-invariant combinations of these two point functions.

\subsubsection{Massless theory in $D = 4$}

The Fierz-Pauli massive graviton propagator is singular in any dimension as the mass goes to zero. In flat space, although massless linearized gravity is not conformal at the level of the action in $D = 4$, the appropriate two-point function of gauge-invariant operators does have the form predicted by conformal symmetry. To see that this is also true also in dS, we first make the substitution $m_{\tilde{h}}^2 = \epsilon H^2$ and then expand the massive graviton two-point function around $\epsilon=0$.  The results for the functions in the propagator \eqref{massgravtwopt2} are 
\begin{align}
f_1(Z) =&\ \frac{H^2}{24 \pi ^2 (Z-1)^3}\bigg[\frac{1}{\epsilon}Z \left(8 - 9 Z + 3 Z^2\right)\nn\\
& +\frac{-2 (Z-1)^3 \left(3Z^2+9 Z+8\right) Z \log \left(\frac{1-Z}{2}\right)-2 Z^6-6 Z^5+5 Z^4+26 Z^3+30 Z^2+4 Z-9}{6\left(Z+1\right)^3}\bigg], \nn\\
f_2(Z) =&\  \frac{H^2}{48 \pi ^2 (Z-1)^3}\bigg[\frac{1}{\epsilon}\left(-9 Z^3+27 Z^2-29 Z+15\right) \nn\\
& +\frac{(Z-1)^3 \left(9Z^3+27 Z^2+29 Z+15\right) \log \left(\frac{1-Z}{2}\right)+3 Z^6+9 Z^5-10 Z^4-34 Z^3+15 Z^2+49 Z+16}{3 \left(Z+1\right)^3}\bigg], \nn\\
f_3(Z) =&\ \frac{H^2}{12 \pi ^2 (Z-1)^3}\bigg[\frac{1}{\epsilon}\left(-3 Z^3+19 Z^2-53 Z+85\right) \label{uglycoeffs}\\
&+\frac{(Z-1)^5 (3 Z+5) \log \left(\frac{1-Z}{2}\right)+(Z+1) \left(Z^5+2 Z^4-22 Z^3+44 Z^2+181Z+82\right)}{3 \left(Z+1\right)^3}\bigg],\nn\\
f_4(Z) =&\ \frac{H^2}{8 \pi ^2 (Z-1)^3 }\bigg[-\frac{1}{\epsilon}(Z+1) \left(Z^2-4 Z+5\right)\nn\\
& +\frac{3 (Z-1)^4 \left(Z^2+4 Z+5\right) \log \left(\frac{1-Z}{2}\right)+Z^6+3 Z^5-18 Z^3-75 Z^2-57 Z+2}{9\left(Z+1\right)^3}\bigg],\nn\\
f_5(Z) =&\ \frac{H^2}{48 \pi ^2 (Z-1)^3}\bigg[\frac{1}{\epsilon}\left(-9 Z^3+37 Z^2-59 Z+55\right)\nn\\
&+\frac{(Z-1)^4 \left(9 Z^2+26 Z+25\right) \log\left(\frac{1-Z}{2}\right)+(Z+1) \left(3 Z^5+6 Z^4-26 Z^3+12 Z^2+63 Z+86\right)}{3\left(Z+1\right)^3}\bigg],\nn
\end{align}
up to terms that vanish as $\epsilon \rightarrow 0$.  The massless singularity now appears as a $1/\epsilon$ simple pole.

From this, we now form the two-point function of the linearized Weyl tensor \cite{Frob:2014fqa,Frob:2014cza}, which is the basic gauge invariant operator, by taking the appropriate derivative combinations,
\begin{align}
W_{\mu\nu\rho\sigma} = -4\mathcal{P}_{\mu\nu\rho\sigma}^{\alpha\beta\gamma\delta} \nabla_\alpha \nabla_\beta h_{\gamma\delta} \,,\label{weylmasllessgee}
\end{align}
where $\mathcal{P}_{\mu\nu\rho\sigma}^{\alpha\beta\gamma\delta}$ is the projection operator onto a tensor with the symmetries of the Weyl tensor, i.e. a fully traceless window tableau $\raisebox{5pt}{\resizebox{.5cm}{!}{\gyoung(\mu\rho ,\nu\sigma )}}$. 
Forming the two-point function of \eqref{weylmasllessgee}, the $1/\epsilon$ poles in \eqref{uglycoeffs} cancel and we find\footnote{As was the case in flat space, getting the correlator into this form requires more than simple tensor manipulations, it requires using dimensionally dependent identities in $D = 4$. We accounted for this by evaluating the expressions explicitly in components.}
\begin{align}
\langle W_{\mu\nu\rho\sigma}(x) W_{\mu'\nu'\rho'\sigma'}(x')\rangle 
=&\ \Omega(x)^{-\Delta + 4}\Omega(x')^{-\Delta + 4} \frac{96}{\pi^2 |x-x'|^{2\Delta}}\mathcal{P}_{\mu\nu\rho\sigma}^{\alpha\beta\gamma\delta} \mathcal{P}_{\mu'\nu'\rho'\sigma'}^{\alpha'\beta'\gamma'\delta'} I_{\alpha\alpha'} I_{\beta\beta'} I_{\gamma\gamma'} I_{\delta\delta'}\,,
\end{align}
with 
\begin{align}
\Omega(x) = \frac{1}{\left(1+\frac{H^2}{4} x^2\right)},\quad{} \Delta = \frac{D}{2} + 1 = 3\,.
\label{omegaDeltaWeyl}
\end{align}
This now has the structure of Weyl factors times the flat space result found in \cite{Farnsworth:2021zgj}, exactly of the form \eqref{weylrelatedcorree} and \eqref{deltatodwe} with $\Delta=3$, $r=4$ and $\Delta_W=-1$ . This means that although massless gravity was not conformal in terms of standard transformations of the Lagrangian \eqref{spin2lagegfpe}, the two-point function of the Weyl tensor (the appropriate gauge-invariant operator for massless linearized gravity) does have the correct conformally invariant structure, as was the case in flat space. In this sense, massless gravity in $D = 4$ on any maximally symmetric background is conformal. 

The massless linear spin-2 theory in flat space in $D>4$ is a unitary scale invariant but not conformally invariant theory.  In \cite{Farnsworth:2021zgj} we saw, following the pattern of what happens in the Maxwell spin-1 case \cite{El-Showk:2011xbs}, that the correlators of the Weyl tensor have the form of a descendant operator, and so the theory can be embedded into a larger non-unitary conformal theory by including a spin-2 primary which violates the unitarity bound.  On (A)dS space, we find that this construction does not work.  For example, in $D=6$, we checked that taking a descendent of the conformal structure \eqref{hhconf} does not yield the massless correlator, $\mathcal{P}\mathcal{P} \langle \nabla_{\rho}\nabla_{\sigma}\bar{h}_{\mu\nu}(x) \nabla_{\rho'}\nabla_{\sigma'}\bar{h}_{\mu'\nu'}(x')\rangle \not= \langle W_{\mu\nu\rho\sigma}(x) W_{\mu'\nu'\rho'\sigma'}(x')\rangle$, though they agree in the flat limit.

\subsubsection{Conformal partially massless theory in $D = 4$}

In $D=4$, the conformal correlators \eqref{trtensorxeceae}, \eqref{finalcorrdn4ee} diverge due to the additional PM gauge symmetry in the conformal theory.  We will see in this section that this divergence cancels when we compute the correlators of gauge-invariant operators.  The basic gauge-invariant operator in this case is the tensor $C_{\mu\nu\rho}$ defined in \eqref{Cdefinitione}.

In $D=4$ the conformal action diagonalizes to \eqref{confd4lage}. 
We can study correlators of $C_{\rho\mu\nu}$ by looking at the correlators of $\tilde h_{\mu\nu}$ and $A_\mu$ of this diagonal action, then use the field redefinition \eqref{PMfieldredefe} to form the correlators of the original PM field $h_{\mu\nu}$,
\begin{align}
 h_{\mu\nu}&= \tilde{h}_{\mu\nu}+ \nabla_{(\mu}A_{\nu)}\,,
 \end{align}
 and then form the combination \eqref{Cdefinitione}.
  In terms of two-point functions, we have
\begin{align}
 \begin{split}
 &\langle h_{\mu\nu}(x) h_{\mu'\nu'}(x')  \rangle = \langle \tilde h_{\mu\nu}(x) \tilde h_{\mu'\nu'}(x')  \rangle \\
 &\quad +\frac{1}{4}\left(\nabla_\mu \nabla_{\mu'} \langle A_{\nu}(x) A_{\nu'}(x')  \rangle+\nabla_\mu \nabla_{\nu'}\langle A_{\nu}(x) A_{\mu'}(x')  \rangle+\mu\leftrightarrow \nu\right) . 
 \end{split}
 \end{align}
 Here $\langle \tilde h_{\mu\nu}(x) \tilde h_{\mu'\nu'}(x')  \rangle$ is the graviton two-point function \eqref{massgravtwopt2}, and $\langle A_{\mu}(x) A_{\mu'}(x')  \rangle$ is the vector two-point  function \eqref{veccorrgen}. These are both singular in $D= 4$ in the limits of the masses we want, $m_{\tilde{h}}^2 = 2 H^2,\ m_A^2= 0$.  We isolate these divergences by setting $m_{\tilde{h}}^2 = (2+\epsilon) H^2$ and $m_A^2= \epsilon H^2 $ and expanding in $\epsilon$.   The divergences then appear as $1/\epsilon$ poles. 
 
 Forming the correlator of the gauge invariant combination \eqref{Cdefinitione}, the $1/\epsilon$ poles cancel and we are left with the finite two-point function
 \begin{align}
\langle C_{\mu\nu\rho}(x) C_{\mu'\nu'\rho'}(x')\rangle 
= \frac{H^4}{4\pi^2 (1-Z)^2}\mathcal{P}_{\mu\nu\rho}^{\alpha\beta\gamma}\mathcal{P}_{\mu'\nu'\rho'}^{\alpha'\beta'\gamma'} \left( g_{\alpha\alpha'} + n_\alpha n_{\alpha'} \right) \left( g_{\beta\beta'} + n_\beta n_{\beta'} \right) \left( g_{\gamma\gamma'} + n_\gamma n_{\gamma'} \right),
 \end{align}
 where $\mathcal{P}_{\mu\nu\rho}^{\alpha\beta\gamma}$ is the projector onto the traceless hook tableau $\raisebox{5pt}{\resizebox{.5cm}{!}{\gyoung(\mu\rho ,\nu )}}$.  Writing this in our coordinates gives
  \begin{align}
\langle  C_{\mu\nu\rho}(x) C_{\mu'\nu'\rho'}(x')\rangle= \Omega(x)^{-\Delta +3} \Omega(x')^{-\Delta +3} \bigg[\frac{1 }{\pi^2|x-x'|^{2\Delta}}\mathcal{P}_{\mu\nu\rho}^{\alpha\beta\gamma}\mathcal{P}_{\mu'\nu'\rho'}^{\alpha'\beta'\gamma'} I_{\alpha\alpha'} I_{\beta\beta'} I_{\gamma\gamma'} \bigg]
 \end{align}
 with
 \begin{align}
\Omega(x) = \frac{1}{\left(1+\frac{H^2}{4} x^2\right)},\quad{} \Delta = \frac{D}{2} = 2\,. 
\label{omegaDeltaPM}
\end{align}
 This is now manifestly conformal as in \eqref{weylrelatedcorree} and \eqref{deltatodwe}, where here $\Delta = 2$, $r = 3$ and $\Delta_W = -1$.

\section{Conclusions\label{conclusion}}

We have investigated the scale and conformal invariance of free 2-derivative scalar, vector and rank-2 symmetric tensor fields on maximally symmetric backgrounds.  In the non-flat cases, there is no algebraic distinction between scale and conformal invariance on these backgrounds, i.e. there is no subalgebra of the conformal algebra that is strictly larger than the isomorphism algebra but strictly smaller than the full conformal algebra. 

For the massive scalar $\phi$ in $D\geq 3$, the action is conformal in any dimension only for the particular value of the mass $m_\phi^2/H^2 ={D(D-2)/ 4}$, and the correlator is conformal here too.  

For the massive vector in $D\geq 3$, the action is conformal only for the massless value where the field becomes gauge invariant, and only in $D=4$.   In the $D=4$ massless case the correlator of the vector field diverges.   We showed that these divergences cancel in the correlators of the gauge-invariant field strength operators $F_{\mu\nu}$, which take the expected conformally invariant form.  

For $D \neq 4$, there is a unique 2-derivative non-gauge-invariant vector theory with non-canonical kinetic terms which is conformal. The conformal invariance of this theory is also reflected in the two-point functions of the vector.  This theory can be diagonalized into a canonical massive vector $\tilde A_\mu$ and massive scalar $\phi$, with the masses $m_{\tilde{A}}^2/H^2={(D-4)(D-2)/ 4}$, $m_\phi^2/H^2 ={D(D-2)/ 4}$.  The kinetic terms of the vector and scalar have opposite signs in dS and same signs in AdS, but the mass squared of the vector is negative on AdS, so in either case the theory is non-unitary.  In this diagonalized form, conformal symmetry acts in a way that mixes the two fields.

For a spin-2 field with the Fierz-Pauli action \eqref{spin2lagegfpe} in $D\geq 3$, the action never has standard conformal symmetry for any mass.  In the massless case, the correlators of the graviton diverge, but those of the gauge-invariant linearized Weyl tensor $W_{\mu\nu\rho\sigma}$ are finite.   The Weyl tensor correlators become conformally invariant only for $D=4$, analogously to what was demonstrated in flat space in \cite{Farnsworth:2021zgj} (see appendix C of \cite{Benedetti:2023jye} for the spin 3/2 case).  

For a rank 2 symmetric tensor field, although there is no conformal Lagrangian for the usual Fierz-Pauli action, there is one if we allow for non-canonical 2-derivative kinetic terms.  There is a trace part that decouples and is equivalent to the conformal scalar, and once this is removed there is a unique conformal Lagrangian for a traceless tensor.  The conformal invariance of this theory is also reflected in the two-point functions of the traceless tensor.   This theory can be diagonalized into canonical massive actions for a massive spin-2 field $\tilde h_{\mu\nu}$, a massive spin-1 field $A_\mu$, and a massive scalar field $\phi$, with the masses $m^2_{\tilde{h}}/H^2=  {D(D-2)/ 4}$, $m_A^2/H^2={(D-4)(D-2)/ 4}$, $m_\phi^2/H^2 ={D(D-2)/ 4}$.  The kinetic terms of the fields alternate signs in dS and have the same signs in AdS, but the mass squared of the vector and tensor are negative on AdS, so in either case the theory is non-unitary.  In this diagonalized form, conformal symmetry acts in a way that mixes the three fields.   

When $D=4$, this conformal theory gets an enhanced 2-derivative scalar gauge symmetry.  In this case the tensor becomes partially massless, the vector becomes massless, and the scalar disappears.  This leaves a conformally invariant theory of a partially massless spin-2 and a photon, where the conformal symmetry mixes the two.  This theory can be expressed in terms of a rank 3 mixed symmetry Weyl-like tensor $C_{\mu\nu\rho}$, and is in many ways analogous to linearized conformal gravity.  The correlators of the graviton diverge in this theory, but we showed that correlators of $C_{\mu\nu\rho}$ are finite and conformally invariant.  {This is in contrast to the correlators of the gauge invariant operators in the pure PM theory, which are finite but not conformally invariant.}

 {Finally, we saw that electromagnetism and linearized gravity in $D>4$ cannot generally be embedded into a larger non-unitary conformal theory in the way that it was done in \cite{El-Showk:2011xbs,Farnsworth:2021zgj} for flat space.}

There is a natural conjecture that extends these results to arbitrary spin.
 For every $s\geq 1$, there is a two derivative conformal Lagrangian for a traceless field $\tilde h_{\mu_1\cdots \mu_s}$, namely the $\kappa=1$ spin $s$ ``special'' conformal field in the classification of \cite{Metsaev:2016oic}.  Its (A)dS$_D$ extension should factorize into spins $s'=s,s-1,\ldots,0$ with masses 
\be m^2_{(s')}=\begin{cases} 
{D(D-2)\over 4}H^2\,,  & s'=0\, , \\
{(D+2s'-4)(D+2s'-6)\over 4}H^2\,, & s'=1,2,\ldots, s\,.
\end{cases}
\ee
These are the masses that correspond to the dual CFT dimensions $\delta_\pm={d\pm 1\over 2}$ through the (A)dS/CFT mass formula
\be \delta_\pm = \begin{cases}  {d\over 2}\pm\sqrt{{d^2\over 4}-{m^2\over H^2}}\, , & s= 0 \, , \\  {d\over 2}\pm\sqrt{{\left(d+2(s-2)\right)^2\over 4}-{m^2\over H^2}} \, , & s\geq 1\, , \label{gensadscctm}
\end{cases}
\ee
where $d \equiv D-1$.  The relative signs in front of the kinetic terms in the factorized Lagrangian should alternate on dS and be the same on AdS.  Conformal symmetry should act linearly but mix all the fields into each other.  For $D=4$, the original conformal field acquires a PM-like gauge symmetry $\delta \tilde h_{\mu_1\cdots \mu_s}= \nabla _{(\mu_1}\cdots \nabla_{\mu_s)_T}\Omega$ with a scalar gauge parameter $\Omega$ (for $s\geq 2$ these are the $\kappa=1$ case of ``type II partial-short'' in \cite{Metsaev:2016oic}, and for $s=1$ it is called ``short'').   This gauge symmetry should act to remove the scalar longitudinal modes from all the component fields: indeed when $D=4$ the component fields $s'\geq 1$ all take the mass values of the maximal depth PM field (i.e. the one with a scalar gauge parameter), and the scalar mode $s'=0$ should be absent.  We can see that this multiplet satisfies the condition of \cite{Barnich:2015tma} for a gauge theory to be conformal: the reducibility parameter for a spin $s'$ maximal depth PM field forms a rank $s'-1$ traceless symmetric tensor of the isometry algebra $\frak{so}(1,4)$.  Taken together, the reducibility parameters for all the $s'=s,s-1,\ldots,1$ combine into a rank $s-1$ symmetric traceless tensor of the conformal algebra $\frak{so}(2,4)$.   All of this should be reflected in the correlators, in a manner analogous to what we have worked out here for spins $\leq 2$.

\vspace{-.2cm}
\paragraph{Acknowledgments:} We would like to thank Nicolas Boulanger, Sylvain Thom\'ee and Matt Walters for helpful conversations and comments.  The work of KF is supported by the Swiss National Science Foundation under grant no. 200021-205016.  KH acknowledges support from DOE grant DE-SC0009946. 

\appendix
\section{Maximally symmetric bi-tensors\label{appendixbi-tensor}}

Here we review the bi-tensor formalism used to construct correlators on maximally symmetric spaces \cite{Allen:1985wd}.  

We are interested in the case of Euclidean dS, which is the sphere of radius $1/H$.  The 2 point correlators depend on two generic points $x$ and $x'$ on the sphere.  Let $\mu$ be the minimal geodesic distance between $x$ and $x'$.  It is convenient to use the variable $Z$ defined as
\be Z\equiv\cos(H\mu),\ \ \ \mu={1\over H}\cos^{-1}Z\,.\label{Zdefie}\ee
$Z\in [-1,1]$ and $\mu\in [0,\pi/H]$, with $Z=1$, $\mu=0$ the coincident point configuration and $Z=-1$, $\mu=\pi/H$ the antipodal configuration on the sphere.

Let $g_{\mu\nu}(x)$ be the metric at $x$ and $g_{\mu'\nu'}(x')$ be the metric at $x'$.  Let $n_\mu(x,x')$ be the unit normal vector at $x$, pointing away from $x'$ along the minimal geodesic connecting $x$ to $x'$.  We have
\be n_\mu=\nabla_\mu \mu=-{1\over H}{1\over \sqrt{1-Z^2}}\nabla_\mu Z\,,\ \ \ \nabla_\mu Z= -H \sqrt{1-Z^2} n_\mu ,\ \ \  g_{\mu\nu}n^\mu n^\nu=1 \,. \ee
Let $n_{\mu'}(x,x')$ be the unit normal vector at $x'$, pointing away from $x$ along the minimal geodesic connecting $x$ to $x'$.  We have
\be n_{\mu'}=\nabla_{\mu'} \mu=-{1\over H}{1\over \sqrt{1-Z^2}}\nabla_{\mu'} Z\,,\ \ \ \nabla_{\mu'} Z= -H \sqrt{1-Z^2}  n_{\mu'} \,,\ \ \  g_{\mu'\nu'}n^{\mu'} n^{\nu'}=1\,. \ee
Let $g_{\mu\mu'}(x,x')$ be the operator that parallel transports indices between $x'$ and $x$.   We have
\be  n_\mu= -g_{\mu\mu'}n^{\mu'} \, ,\ \ \ n_{\mu'} = -n^{\mu}   g_{\mu\mu'}\,.\label{nparalleltee}\ee

We can compute the covariant derivatives of these functions following \cite{Allen:1985wd} and we get:
\be \nabla_\mu n_\nu=H{Z\over \sqrt{1-Z^2}}\left( g_{\mu\nu}-n_\mu n_\nu\right),\ \ \  \nabla_{\mu'} n_{\nu'}=H{Z\over \sqrt{1-Z^2}}\left( g_{\mu'\nu'}-n_{\mu'} n_{\nu'}\right), \nn\ee
\be \nabla_\mu n_{\nu'}=-H{1\over \sqrt{1-Z^2}}\left( g_{\mu\nu'}+n_\mu n_{\nu'}\right),\ \ \   \nabla_{\mu'} n_{\nu}=-H{1\over \sqrt{1-Z^2}}\left( g_{\nu\mu'}+n_{\mu'}n_\nu \right), \nn\ee
\be \nabla_\mu g_{\nu\rho'}=H{1-Z\over \sqrt{1-Z^2}}\left( g_{\mu\nu}n_{\rho'}+ g_{\mu\rho'}n_{\nu}\right), \ \ \ \nabla_{\mu'} g_{\rho\nu'}=H{1-Z\over \sqrt{1-Z^2}}\left( g_{\mu'\nu'}n_{\rho}+ g_{\rho\mu'}n_{\nu'}\right).  \label{ngnderivresee}\ee

The objects $n_\mu$, $n_{\mu'}$, $g_{\mu\mu'}$ are parallel transported along the geodesic and thus 
\be n^\mu \nabla_\mu n_{\nu} =n^\mu \nabla_\mu n_{\nu'}  =n^\mu \nabla_\mu g_{\nu\nu'}=0, \label{paraleltranse1e}\ee  
\be n^{\mu'} \nabla_{\mu'} n_{\nu} =n^{\mu'} \nabla_{\mu'} n_{\nu'}  =n^{\mu'} \nabla_{\mu'} g_{\nu\nu'}=0\,, \ee  
as can be verified from \eqref{ngnderivresee}.

We will often use conformally flat coordinates on the sphere, 
\be g_{\mu\nu}(x)=\Omega^2(x)\eta_{\mu\nu}\, ,\ \ \ \Omega(x)={1\over 1+{H^2\over 4}x^2}\, ,\ee
in which we have
\be Z=1-{H^2\over 2} \Omega(x)\Omega(x') (x-x')^2.\label{conformalepforZe}\ee
In these coordinates, the geometric quantities above all manifestly reproduce the flat space limit as $H\rightarrow 0$, 
\begin{align}
 n_\mu(x,x')&={H \Omega(x)\over 2\sqrt{1-Z^2}}\left( 2\Omega(x') (x - x')_\mu -(1-Z) x_\mu\right)={(x-x')_\mu\over |x-x'|}+{\cal O}(H^2)\, ,\\
 n_{\mu'}(x,x')&=-{ H \Omega(x')\over 2\sqrt{1-Z^2}}\left( 2\Omega(x) (x - x')_{\mu'} +(1-Z) x_{\mu'}\right)=-{(x-x')_{\mu'}\over |x-x'|}+{\cal O}(H^2)\, ,
 \end{align}
\begin{align}
\begin{split}
 g_{\mu\mu'}(x,x') &= \Omega(x) \Omega(x') \eta_{\mu\mu'} -2 n_\mu n_{\mu'}- H^2\Omega^2(x) \Omega^2(x') \frac{(x-x')_\mu (x-x')_{\mu'}}{(1-Z)} \\
&= \eta_{\mu\mu'}+{\cal O}(H^2)\, .
\end{split}
\end{align}

In these coordinates, the standard conformal tensor structure that appears in 2-pt. correlators also takes a simple form,
\begin{align}
\begin{split}
 g_{\mu\mu'}+2n_\mu n_{\mu'}&=\Omega(x) \Omega(x') \left[\eta_{\mu\mu'} - 2 \frac{(x-x')_\mu (x-x')_{\mu'}}{(x-x')^2}\right]\\
 &=\Omega(x) \Omega(x') I_{\mu\mu'}\,, 
\end{split}
\label{conftensorcoords}
\end{align}
where $I_{\mu\mu'}\equiv \eta_{\mu\mu'} - 2 \frac{(x-x')_\mu (x-x')_{\mu'}}{(x-x')^2}$ is the flat space conformal tensor structure.

\newpage

\renewcommand{\em}{}
\bibliographystyle{utphys}
\addcontentsline{toc}{section}{References}
\bibliography{dSconformal-arxiv-v2}

\end{document}